\documentclass[12pt]{iopart}
\usepackage{iopams}
\expandafter\let\csname equation*\endcsname\relax
\expandafter\let\csname endequation*\endcsname\relax
\usepackage{amsmath}

\usepackage{graphicx}
\usepackage{hyperref}
\usepackage{verbatim}
\usepackage{array}
\usepackage[table]{xcolor}
\usepackage{multirow}
\usepackage[section]{placeins} %This prevents placing floats before a section.
\usepackage[square,numbers,comma,sort&compress]{natbib} % bibliography

\usepackage{comment}

\usepackage{tabularx,ragged2e}
\newcolumntype{Y}{>{\RaggedRight\arraybackslash\hspace{0pt}}X}
\newcolumntype{C}{>{\Centering\arraybackslash\hspace{0pt}}X}

\usepackage[T1]{fontenc}

\graphicspath{{images/}{images/nema_image_quality/}{images/nema_sensitivity/}{images/nema_sf_necr/}{images/nema_spatial_resolution/}}

\begin{document}

\title[Simulating NEMA characteristics of the modular total-body J-PET scanner - an economic total-body PET from plastic scintillators]
{Simulating NEMA characteristics of the modular total-body J-PET scanner - an economic total-body PET from plastic scintillators}

\author{
P.~Moskal$^{1,2}$,
P.~Kowalski$^3$,
R.~Y.~Shopa$^3$,
L.~Raczy\'nski$^3$,
J.~Baran$^4$,
N.~Chug$^{1,2}$,
C.~Curceanu$^5$,
E.~Czerwi\'nski$^{1,2}$,
M.~Dadgar$^{1,2}$,
K.~Dulski$^{1,2}$,
A.~Gajos$^{1,2}$,
B.~C.~Hiesmayr$^6$,
K.~Kacprzak$^{1,2}$,
\L.~Kap\l on$^{1,2}$,
D.~Kisielewska$^{1,2}$,
K.~Klimaszewski$^3$,
P. Kopka$^3$,
G.~Korcyl$^{1,2}$,
N.~Krawczyk$^{1,2}$,
W.~Krzemie\'n$^7$,
E.~Kubicz$^{1,2}$,
Sz.~Nied\'zwiecki$^{1,2}$,
Sz.~Parzych$^{1,2}$,
J.~Raj$^{1,2}$,
S.~Sharma$^{1,2}$,
S.~Shivani$^{1,2}$,
E.~St\c{e}pie\'n$^{1,2}$,
F.~Tayefi$^{1,2}$,
W.~Wi\'slicki$^3$
}

\address{$^1$ Faculty of Physics, Astronomy and Applied Computer Science, Jagiellonian University, 30-348 Cracow, Poland}
\address{$^2$ Total-Body Jagiellonian-PET Laboratory, Jagiellonian University, 30-348 Cracow, Poland}
\address{$^3$ Department of Complex Systems, National Centre for Nuclear Research, 05-400 Otwock-\'Swierk, Poland}
\address{$^4$ Institute of Nuclear Physics Polish Academy of Sciences, 31-342 Cracow, Poland}
\address{$^5$ INFN, Laboratori Nazionali di Frascati, 00044 Frascati, Italy}
\address{$^6$ Faculty of Physics, University of Vienna, 1090 Vienna, Austria}
\address{$^7$ High Energy Physics Division, National Centre for Nuclear Research, 05-400 Otwock-\'Swierk, Poland}

\begin{abstract}

The purpose of the presented research is estimation of the performance characteristics of the economic Total-Body Jagiellonian-PET system (TB-J-PET) constructed from plastic scintillators.
The characteristics are estimated according to the NEMA NU-2-2018 standards utilizing the GATE package.
The simulated detector consists of 24 modules, each built out of 32 plastic scintillator strips (each with cross section of 6 mm times 30 mm and length of 140 cm or 200 cm) arranged in two layers in regular 24-sided polygon circumscribing a~circle with the diameter of 78.6~cm.

For the TB-J-PET with an axial field-of-view (AFOV) of 200 cm, a~spatial resolutions of 3.7 mm (transversal) and 4.9 mm (axial) are achieved.
The noise equivalent count rate (NECR) peak of 630 kcps is expected at 30 kBq/cc activity concentration and the sensitivity at the center amounts to 38 cps/kBq.
The scatter fraction is estimated to 36.2~\%.
The values of scatter fraction and spatial resolution are comparable to those obtained for the state-of-the-art clinical PET scanners and the first total-body tomographs: uExplorer and PennPET.
With respect to the standard PET systems with AFOV in the range from 16~cm to 26~cm, the TB-J-PET is characterized by an increase in NECR approximately by factor of 4 and by the increase of the whole-body sensitivity by factor of 12.6 to 38.
The time-of-flight resolution for the TB-J-PET is expected to be at the level of CRT~=~240~ps (FWHM).
For the TB-J-PET with an axial field-of-view (AFOV) of 140 cm, an image quality of the reconstructed images of a NEMA IEC phantom was presented with a contrast recovery coefficient (CRC) and a background variability parameters.

The increase of the whole-body sensitivity and NECR estimated for the TB-J-PET with respect to current commercial PET systems makes the TB-J-PET a promising cost-effective solution for the broad clinical applications of total-body PET scanners.
TB-J-PET may constitute an economic alternative for the crystal TB-PET scanners, since plastic scintillators are much cheaper than BGO or LYSO crystals and axial arrangement of the strips significantly reduces the costs of readout electronics and SiPMs.

\end{abstract}

\vspace{2pc}
\submitto{\PMB}
\maketitle

%==========================================================
\section{Introduction}
\label{sec::introduction}
%==========================================================

Positron emission tomography (PET) is a~well established diagnostic method enabling detection of a~tissue pathology on a~molecular level before it evolves to the functional or morphological abnormalities~\cite{schmall2019,mckenney2018}.
Currently, routine PET imaging with devices of about 20~cm AFOV~\cite{grant2016,vansluis2019}, in a~single bed position, enables the diagnosis of individual organs only, and the diagnosis of the whole-body requires a~combination of series of sequential images obtained from many patient positions in the scanner, thus only a~time-dependent scan of the total-body is available~\cite{houshmand2015}.
With the advent of the total-body PET (TB-PET), precision medicine will be enhanced with a~new toolbox that allows for the simultaneous molecular imaging of the whole human body, providing concurrent imaging of metabolic rate in near and distant organs ~\cite{VandenbergheMoskal2020,Moskal2020clinical,Badawi2019,karp2020,Cherry2017,Cherry2018,zhang2020,surti2020,surti2020total,jones2020total,efthimiou2020new}.
Thanks to the high sensitivity, TB-PET enables the extreme reduction of the whole-body imaging duration or the lessen of the radiopharmaceutical dose~\cite{Majewski2020}, thus opening perspectives for application of PET to the wider group of patients (e.g. children~\cite{nardo2020} or patients suffering from systemic diseases~\cite{mckenney2018,nakajima2017,yamashita2014,VandenbergheMoskal2020,borja2020update}).
By the introduction of uExplorer, the first total-body PET, to the clinical practice~\cite{Badawi2019} it was demonstrated~\cite{zhang2020,zhang2019} that in addition to the static standardized uptake value (SUV) images, TB-PET may also deliver a~kinetic model based parametric imaging of all tissues in the body, simultaneously.
These new capabilities open promising prospects for quantitative improvements of diagnostic and prognostic assessments of e.g. oncological, cardiological and neurological diseases~\cite{Majewski2020}.

Yet, the high costs of the TB-PET scanner based on LYSO crystal scintillators, estimated to around $\$$10 million or more~\cite{Cherry2018}, is a~serious obstacle to the widespread use of this modality in clinical practice, including the medical research clinics.
Therefore, reducing the cost of TB-PET production has become one of the important challenges taken by many research groups, and constitutes one of the hot research topics in this field.
The considered solutions include: reduction of scintillator thickness~\cite{surti2013,surti2020total}, reduction of number of detectors by arranging them in sparse configuration~\cite{zhang2019sparse,zein2020physical,zhang2017system} or taking advantage of the Cherenkov light to improve timing properties with BGO crystals~\cite{zhang2017system,gonzalez2017highly,brunner2017bgo,kwon2019dual,cates2019,gundacker2020experimental}.
In order to take advantage of the Compton scattering there is also an ongoing research aiming at combining standard PET tomography with Compton cameras \cite{Grignon2007,Donnard2012,Lang2012,Lang2014,Oger2012,Thirolf2015,Yoshida2020,kuramoto2017development,shimazoe2020development,hemmati2017compton,kishimoto2017first,uenomachi2020double}.

LYSO crystals account for about 50\% of the total costs of the TB-PET scanner, while the rest of the costs comes mainly from readout electronics and SiPMs.
An application of BGO crystals may reduce costs of the scintillators only by a~factor of about~2~\cite{VandenbergheMoskal2020}.
Thus, as it was argued in reference~\cite{Moskal2020clinical}, that the reduction of crystal thickness or exchange of the LYSO by BGO crystals will not lead to a~sufficiently significant reduction of production costs.
%Total cost of components of the plastic TB-PET is about 5 times less with respect to the crystal based TB-PET systems. Cost of crystal scintillators constitutes about 50\% of the total costs of the TB-PET scanner.
Therefore, one of the promising, economic solution for the construction of the TB-PET scanner is the exchange of the expensive crystals to cost-effective plastic scintillator strips arranged axially, further on referred to as Jagiellonian Positron Emission Tomography (J-PET) ~\cite{Moskal2020clinical,Moskal2011,Moskal:2014sra,Moskal:2014rja,Moskal:2016ztv,Szymon-Acta,sharma2020hit,sharma2020estimating}.
The costs of components for the total-body plastic J-PET are expected to be about 5~times less than the crystal-based total-body PET~\cite{Moskal2020clinical}.

Table~\ref{tab1} compares basic properties of plastic scintillator with BGO and LYSO crystals which are currently used in PET detectors.
The much higher density of crystals with respect to plastics, and the more effective registration of annihilation photons, can be compensated by the multi-layer geometry~\cite{Moskal:2016ztv} possible with the axial arrangement of plastic strips with the readout at the ends~\cite{Moskal:2014sra}.
The significantly lower light attenuation in plastic with respect to crystals (e.g. plastic scintillator BC-480 is characterized by 7 and 18 times longer light attenuation length compared to BGO and LYSO crystals, respectively) enables the effective light transport even up to 2~m long plastic strips.
Moreover the negligible fraction of photo-electric effect for the interaction of 511~keV photons in plastic scintillators does not preclude the possibility of the scatter fraction reduction.
When using plastic scintillators the scattering in the patient may be suppressed based on the measurement of the energy deposition due to the Compton interaction~\cite{Moskal:2014sra,Moskal:2016ztv}.

\begin{table}[ph]
  \setlength\tabcolsep{3pt} % default value: 6pt
  \footnotesize
  \centering
    \begin{tabularx}{\textwidth}{|c|*{7}{Y|}}
      \hline
      Scintillator & Density [g/cm$^3$] & Light output [photons/MeV] & Decay time [ns] & Fraction of photoelectric effect [\%] & Light attenuation length [cm] & Linear absorption coefficient for 511 keV photons [cm$^{-1}$] \\
      \hline
        LYSO &
        7.1 - 7.4~\cite{mao2013crystal} &
        33200~\cite{saintgobain_lyso} &
        36~\cite{saintgobain_lyso} &
        31 - 32~\cite{nist} &
        21 - 40~\cite{saintgobain,vilardi2006optimization,mao2008optical} &
        0.82 - 0.87~\cite{nist} \\
      \hline
        BGO &
        7.13~\cite{saintgobain_bgo} &
        8000-10000~\cite{saintgobain_bgo} &
        300~\cite{saintgobain_bgo} &
        41~\cite{nist} & 
        55~\cite{chen2004large} &
        0.96~\cite{nist} \\
      \hline
        BC-408/EJ-200 &
        1.023~\cite{saintgobain} &
        10000~\cite{saintgobain_bc} &
        2.1~\cite{saintgobain} &
        6.3 10$^{-5}$~\cite{nist} &
        380~\cite{saintgobain} &
        0.096~\cite{nist} \\
      \hline
    \end{tabularx}
\caption{
Basic properties of LYSO crystal, BGO crystal and BC-408 (equivalent of EJ-200) plastic scintillators important for the design of the PET systems.
The values of attenuation coefficients and the fractions of photoelectric effect were extracted from the data base maintained by the National Institute of Standards and Technology~\cite{nist}.
\label{tab1}
}
\end{table}

Total-body J-PET (TB-J-PET) may constitute an economic alternative for the crystal TB-PET scanners, since plastic scintillators are more than an order of magnitude less expensive than BGO crystals, and plastic PET with axially arranged scintillator strips reduces also significantly costs of readout electronics and SiPMs.
The reduction of the electronics cost may be achieved due to the fact that the readout (except the WLS strips) is placed at the ends of the cylindrical detector compared to the coverage of the cylinder surface in case of the radially arranged blocks of crystal PET detectors~\cite{Moskal2020clinical}.
Prospects and clinical perspectives of TB-J-PET imaging using plastic scintillators were recently described in Refs~\cite{Moskal2020clinical,moskal2019positronium,moskal2019feasibility}.
Prospects for fundamental physical questions can be found in Refs \cite{moskal2016potential,gajos2020sensitivity,hiesmayr2019witnessing,moskal2018feasibility,kaminska2016feasibility}.
In this article we assess the performance characteristics of TB-J-PET constructed from plastic scintillator strips. The spatial resolution (SR), sensitivity (S), scatter fraction (SF), noise equivalent count rate (NECR) and image quality (IQ) for the TB-J-PET are estimated according to the National Electrical Manufacturers Association NEMA NU 2-2018 standards~\cite{NEMA:2018} by using Geant4 Application for Tomographic Emission (GATE)~\cite{Jan2004,Jan2011,sarrut2014review,sarrut2021review} and the dedicated analysis software developed by the J-PET group~\cite{Kowalski2018,Krzemien2020}.
We assess NEMA characteristics for few TB-J-PET configurations.
Transaxial field-of-view with diameter of 78.6~cm is assumed for all studied geometries, while AFOV of 200~cm (total-body) and 140~cm (head and torso) is considered.
In addition, the influence on the result due to the unknown depth-of-interaction (DOI) is studied comparing results obtained assuming: (i) an ideal case that the true interaction point is known, (ii) a~case when the DOI is not known, and (iii) assuming the resolution for the determination of DOI (full width at half maximum - FWHM) is equal to FWHM(DOI)~=~10~mm.
Moreover studies of spatial resolution as a function of maximum allowed axial distance between the interaction points (the maximum accepted oblique angle of the line-of-response (LOR)) were also conducted.

In the section \textit{Materials and Methods} (Sec.~\ref{sec::materials}) we define the geometry and structure of the TB-J-PET scanner, properties of the materials constituting the detector and the assumed temporal and spatial resolutions.
This section describes also briefly the methods of how the NEMA characteristics are evaluated.
Next, results of simulations and analysis are presented in the section \textit{Results} (Sec.~\ref{sec::results}), which is followed by the \textit{Discussion} (Sec.~\ref{sec::discussion}) section including comparison of TB-J-PET characteristics to the performance characteristics of current clinical solutions and the total-body crystal-based PET scanners, in particular to the total-body uExplorer~\cite{Zhang2017, Cherry2017, Cherry2018, Badawi2019} and PennPET Explorer~\cite{karp2020} systems.
While the NEMA norms were designed for PET systems with small AFOV, they may not be adequate for the total-body PET scanners with AFOV exceeding 70~cm long phantoms and emission sources required by the NEMA NU 2-2018~\cite{NEMA:2018}.
Therefore, in the \textit{Discussion} section the results obtained according to the NEMA norm are compared to the values obtained for the longer sources with lengths of 140~cm and 200~cm. 

%==========================================================
\section{Materials and Methods}
\label{sec::materials}
%==========================================================

Plastic scintillators (1.02 – 1.06 g/cm$^3$)~\cite{saintgobain,eljentechnology} are about 7 times less dense than LYSO crystals (7.0 – 7.4 g/cm$^3$)~\cite{mao2013crystal} and the linear attenuation coefficient for the 511~keV photons is much higher for LYSO and BGO ($\mu~=~0.83$~cm$^{-1}$ and $\mu~=~0.96$~cm$^{-1}$, respectively) than for plastic scintillator ($\mu~=~0.096$~cm$^{-1}$)~(see Tab.~\ref{tab1}).
This implies that in a single scintillator layer with thickness of 2 to 3~cm (typical thickness of scintillators used in PET~\cite{Conti2009,VandenbergheMoskal2020}) most of annihilation photons hitting the detector are interacting in case of LYSO and BGO (81\% to 92\% and 85\% to  94\%, respectively) but only about a~quarter (17\% to 25\%)  in case of plastic scintillators.
Therefore, in case of TB-J-PET the registration efficiency is increased significantly by the application of multi-layer detection system with an effective total thickness larger than 3~cm in the case of two layers, see Fig.~\ref{fig::djpet2m}, e.g. for 2 3-cm-thick layers the registration efficiency would be about 44\%.
% 0.17+0.17*0.83 ~= 0.31
% 0.25+0.25*0.75 ~= 0.4375

\subsection{Detector System Configuration}

\begin{figure}[h]
\centering
\includegraphics[height=0.45\textwidth]{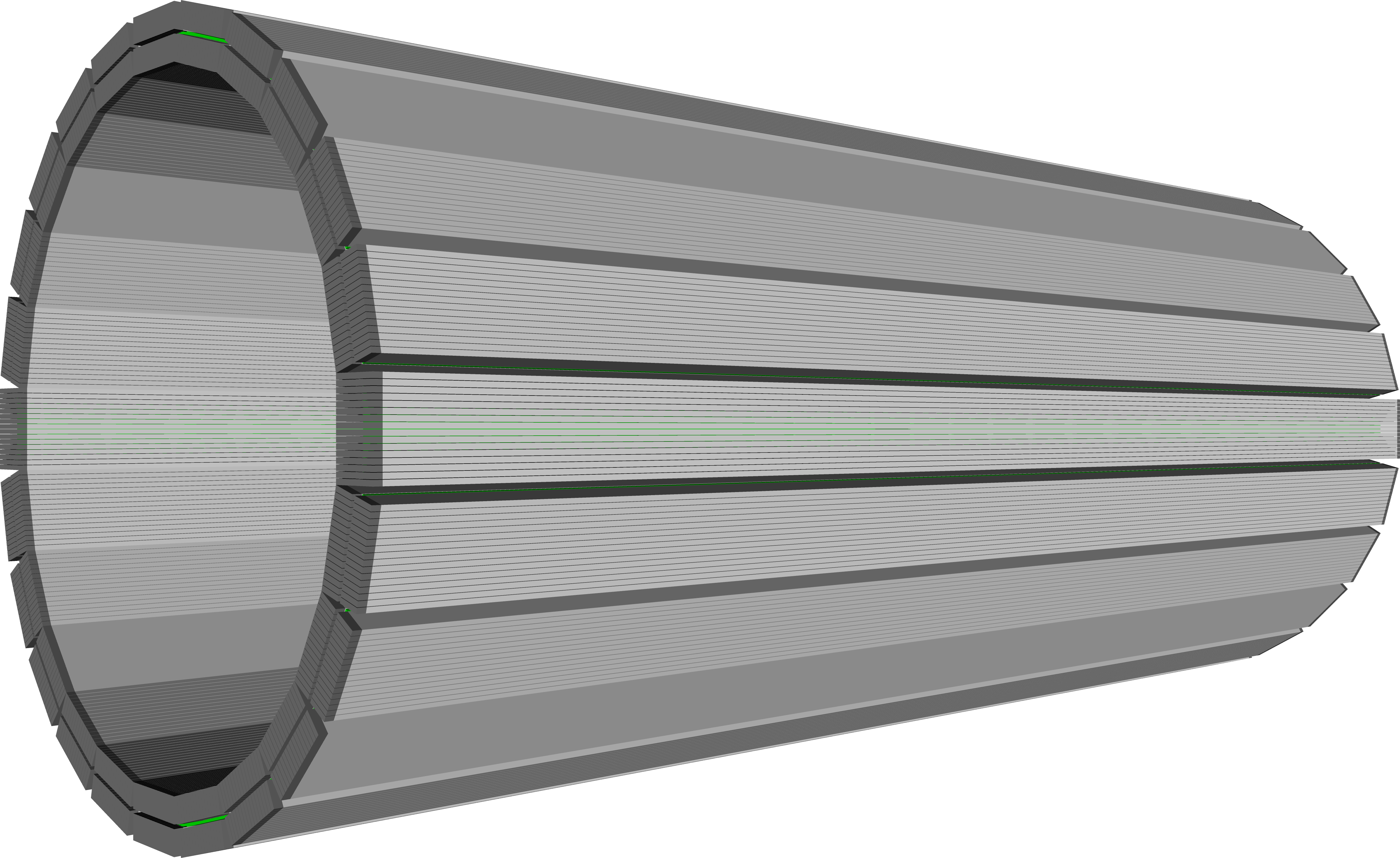}
\includegraphics[height=0.35\textwidth]{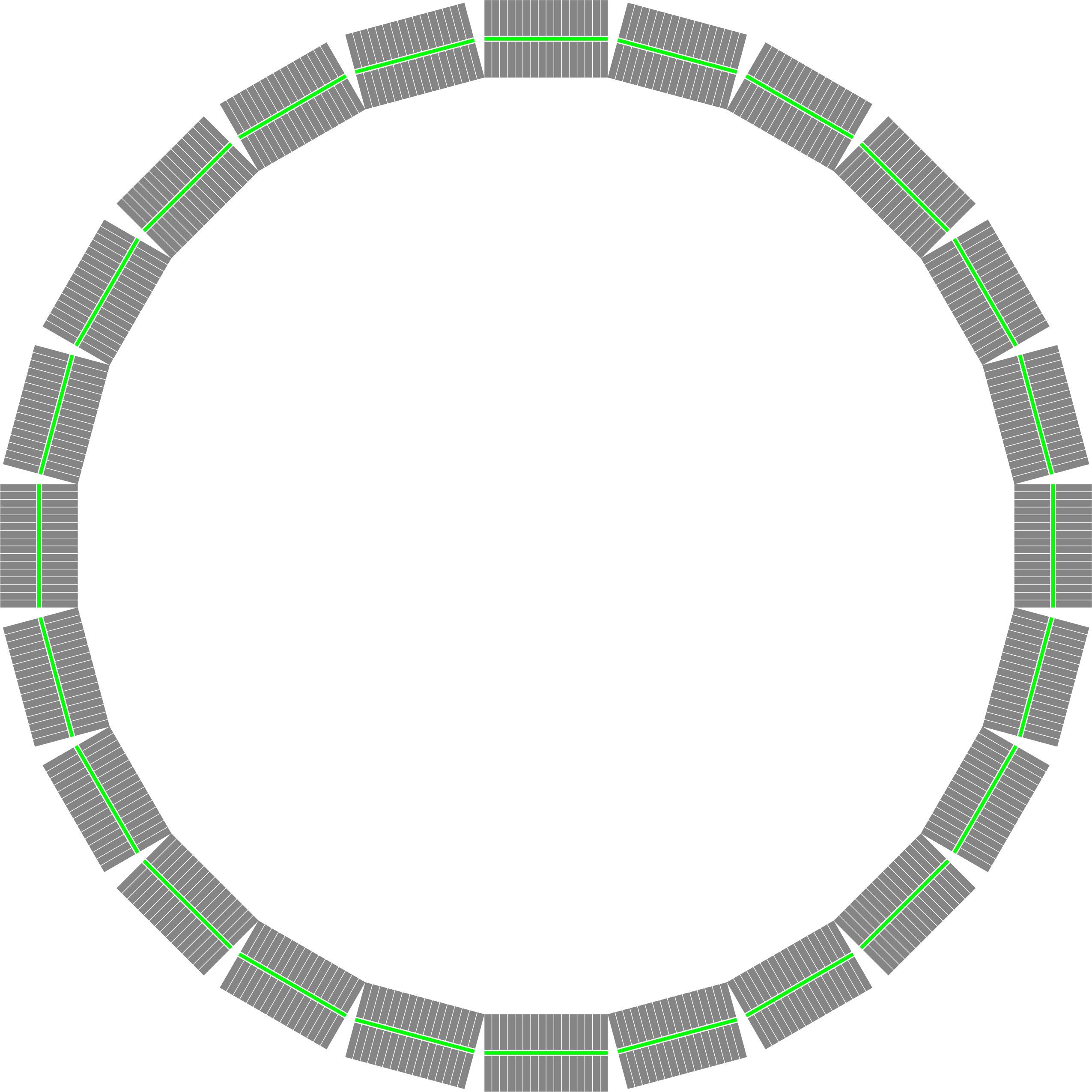}
\includegraphics[height=0.35\textwidth]{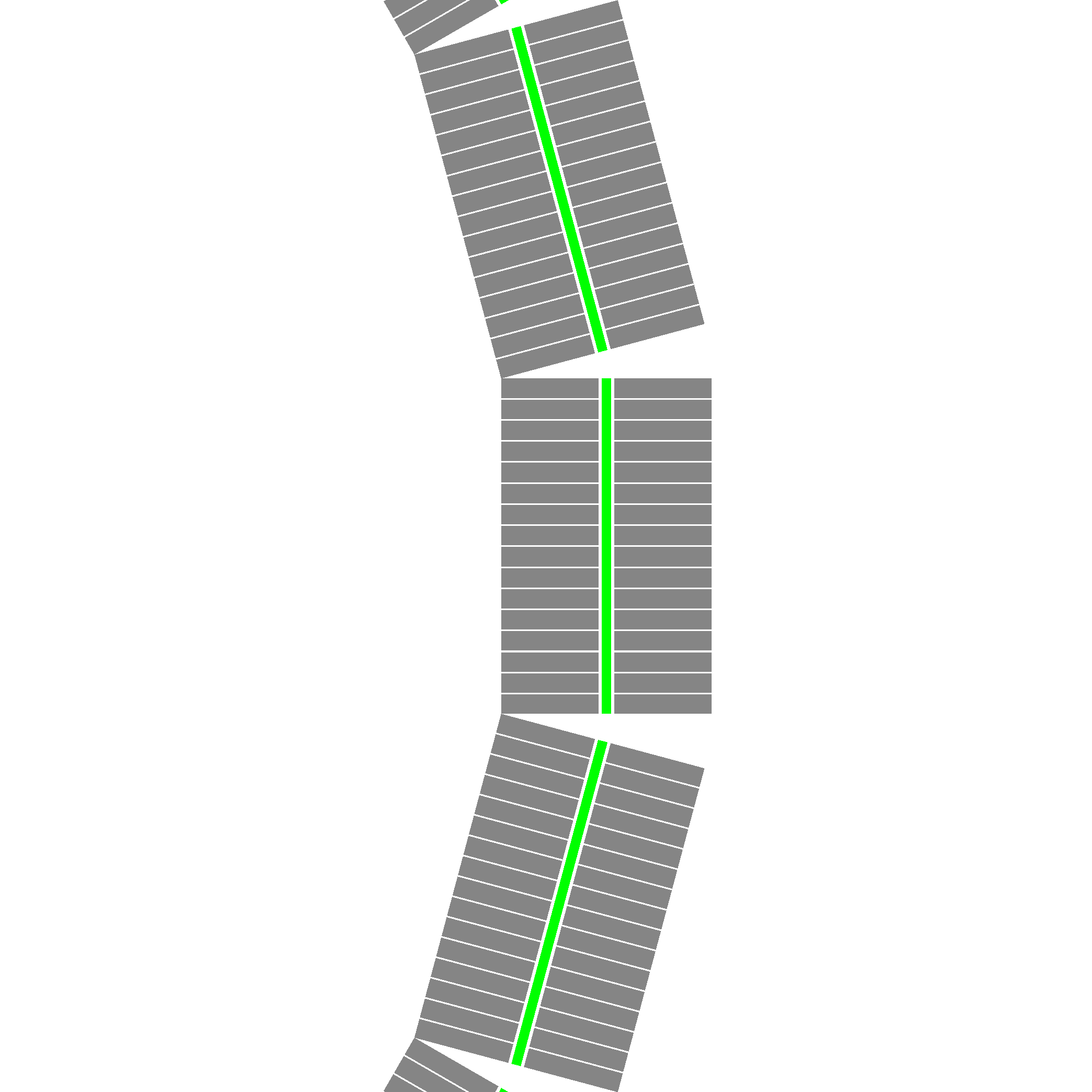}
\caption{
Visualisation of the simulated 2-layer 24-module 2-m-long TB-J-PET scanner.
Scintillator strips are marked in gray and  WLS strips in green.
Upper panel indicates the perspective view of the TB-J-PET scanner and lower panel shows the transverse cross sections: full (left) and zoomed (right). 
}
\label{fig::djpet2m}
\end{figure}

In this article we consider a design of TB-J-PET with a~double layer geometry as it is presented in the upper panel of Fig.~\ref{fig::djpet2m}.
The detector consists of 24 modules, each including 32 scintillator strips arranged in two layers with the additional 3~mm thick layer of wavelength shifters (WLS) (lower panel of Fig.~\ref{fig::djpet2m}).
WLS layer is used for the reconstruction of the axial coordinate of the annihilation photon's interaction point~\cite{smyrski2014application,Smyrski2017,shivani2020development}.
Scintillator strips are rectangular in cross section with dimensions of 0.6~cm~$\times$~3~cm.
In this article, two cases are considered: strips with the length of L~=~140~cm and L~=~200~cm.
The modules are arranged in regular 24-sided polygon circumscribing a~circle with the diameter of D~=~78.6~cm.

The diameter and length of simulated geometries were set to be close to the one chosen for the total-body scanners: uExplorer (L~=~194~cm, D~=~78.6 cm)~\cite{Badawi2019,spencer2020performance} and PennPET Explorer (L~=~140~cm, D~=~81~cm)~\cite{karp2020}.
%Additionally, the diameter of about 80~cm is a~typical diameter for PET devices~\cite{Slomka2016,Vandenberghe2016}.
%in order to simplify comparison with the competitive solution.
%and it is justified with the size of a~typical patient.
%Explorer project is the tomograph with the AFOV of 194 cm and diameter of 78.6 cm with patient bore 76 cm \cite{Badawi2019}.
%The PennPET device is 140~cm AFOV PET device developed by a~team from University of Pennsylvania lead by Professor Joel Karp.
%The diameter of the PennPET scanner is 81 cm.
%2018:"Currently used crystalline tomographs have crystals with depths from 20 mm up to 30 mm, while their cross sections (perpendicular to the radius of the scanner) range from 4 × 4 mm2 to 6.3 × 6.3 mm2
%(Slomka et al 2016, Vandenberghe et al 2016)".
%OPIS J-PETA
%-- moduly
%-- rozmiary 
%-- po co WLSY (Bio, NIM, Shivani, etc..)

\subsection{Spatial and Temporal Resolution}

In order to take into account the detector spatial and temporal resolution the simulated time and axial position of photons' interactions were smeared according to a~Gaussian distribution.
In case of axial resolution the FWHM of 5~mm of the Gaussian was assumed.
Such a~resolution is expected when applying a~layer of WLS strips arranged perpendicularly to the scintillator strips~\cite{smyrski2014application, Smyrski2017, Kowalski2018}.

For the estimation of the influence of the uncertainty of DOI reconstruction three scenarios are considered: (i) ideal case with DOI known from the simulations, (ii) standard case when DOI is estimated as center of the scintillator in the transaxial cross section, and (iii) assuming that the resolution of the determination of DOI can be approximated by the Gaussian function with a~FWHM~=~10~mm.

\begin{figure}[h]
\centering
\includegraphics[width=0.7\textwidth]{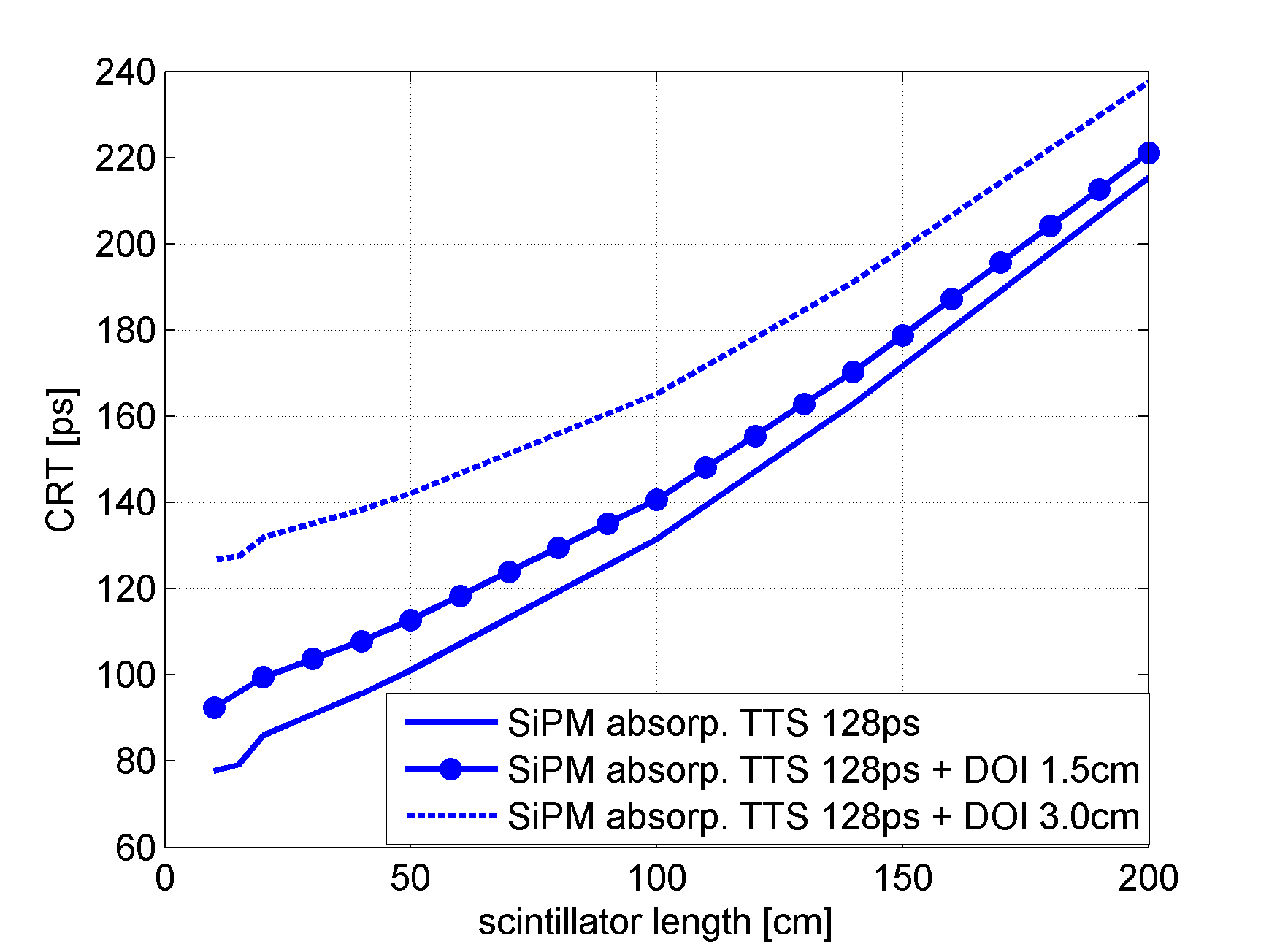}
\caption{
CRT values as a function of the AFOV expected while using BC-408 scintillator.
Values obtained for the radial thickness of 1.5~cm and  3.0~cm, in the case of unknown DOI, are indicated by a solid line with dots and a dotted line, respectively.
The result for ideal case with a known DOI is shown by a solid line. 
}
\label{fig::crts}
\end{figure}

The expected values of coincidence resolving time (CRT) expressed as FWHM were estimated using a~simulation method described in reference~\cite{Moskal:2016ztv} and assuming that detector is built from BC-408 scintillators with a~photon absorption length of 380~cm~(Tab.~\ref{tab1}).
Fig.~\ref{fig::crts} presents the dependence of expected CRT value as a function of the AFOV for the case of unknown DOI in scintillators with thickness of 1.5~cm (solid line with dots), 3~cm (dotted line) and for the ideal case with known DOI (solid line).
For the estimation presented in this article CRT values obtained for the 3~cm thickness with unknown DOI were used, which represents the worst case scenario.

In order to account for the time resolution of the detector system, a simulated time of annihilation photon's interaction was smeared according to the Gaussian distribution corresponding to CRT~=~190~ps and CRT~=~240~ps for AFOV of 140~cm and 200~cm, respectively.

\subsection{Simulation Tools}

Calculations are preformed using the GATE software~\cite{Jan2004,Jan2011,sarrut2014review,sarrut2021review} which enables generation of back-to-back annihilation photons of the sources as defined by the NEMA norm, and simulation of interaction of these photons in the detector material.
Using the dedicated analysis programs the list-mode set of coincidences is obtained~\cite{Kowalski2018,Kowalski2016,Kowalski2015}.
Finally, simulated times and positions of interactions are smeared according to resolutions discussed in the previous sub-section.
Further on in order to reduce contribution of events with photons scattered in the phantom and in the detector, the simulated events are filtered using criteria based on the correlations between the hit time, hit position and energy deposition of annihilation photons in the detector.
The event selection method is described in detail in Ref.~\cite{Kowalski2018}.

\subsection{Image Reconstruction Method}
\label{subsection::TOF-FBP}

3D bin-mode filter-back-projection (FBP) with re-projection (FBP3DRP), implemented in Software for Tomographic Image Reconstruction (STIR) \cite{stir,khateri2019implementation}, was employed in our earlier works \cite{Shopa2017, Kowalski2018, moskal2020performance}.
It is, however, impractical for 2-layer 2-m-long J-PET due to the limitations of the geometries that the STIR framework supports: both multi-layer and continuous scintillators cannot be defined.
Therefore, as a~temporary solution, positions of scatterings inside scintillator strips (hits) are remapped onto a~virtual single cylindrical layer in transverse plane and onto discrete rings along axial direction.
Such procedure, though, may impose additional distortion and worsen axial resolution, undermining the benefits of re-projection at the same time (see the discussion in \cite{Shopa2017}).

Alternatively, we developed a time-of-flight (TOF) FBP algorithm (TOF FBP), based on the idea presented in \cite{Conti2005}.
Projection data is weighted twice: with a~filter in frequency space and during back projection for each TOF bin.
One could also define a~more general, non-bin definition based on the summation over lines-of-response (LORs).
The reconstructed image $f$ for an arbitrary voxel $\textbf{v}$ is estimated using back-propagation operator $\mathcal{B}$ as:

\begin{equation}
\label{eq:TOFFBP}
f(\textbf{\textit{v}}) = \sum_{i=1}^{N_{\text{LOR}}}f_i(\textbf{\textit{v}}) = \sum_{i=1}^{N_{\text{LOR}}}\mathcal{B}p^F_i,
\end{equation}

\noindent where each filtered projection element

\begin{equation}
\label{eq:TOFpF}
p^F_i\equiv p^F_i(s,\phi,\zeta,\theta,t)=\mathcal{F}^{-1}\lbrace W(\nu_s)\mathcal{F} \left[p_i(s,\phi,\zeta,\theta)\right] \rbrace\cdot h(t-t_i).
\end{equation}

\noindent $s$, $\phi$, $\theta$, $\zeta$ are coordinates in projection domain of sinograms $p_i$ \cite{PETBasicScience}, $\mathcal{F}$ and $\mathcal{F}^{-1}$ are 1D Fourier and inverse Fourier transform operators, respectively, which serve to operate between space domain (projection coordinate $s$) and frequency domain (coordinate $\nu_s$).
$W(\nu_s)$ is the ramp filter, $i$ denotes the index of LOR, $h(t)$ is TOF kernel and $t_i$ is the temporal parameter which defines the position of annihilation along LOR.
For each ($i$-th) coincidence, the dimensionality of $p^F_i$ is reduced from $\mathbb{R}^5$ to $\mathbb{R}^2$, since only $s$ and $t$ are not fixed, i.e. $p^F_i\equiv p^F_i(s,\phi_i,\zeta_i,\theta_i,t)$.

The sum in equation~\ref{eq:TOFFBP} requires multiple integration over every LOR, which could be time consuming.
Besides, TOF normalisation is essential, because each back-projection $\mathcal{B}p^F_i$ is not made over the entire image plane, but is restricted by $h(t)$ kernel instead.
However, this can be ignored for the case of 1-mm NEMA source, since it is definitely smaller than the minimal sampling for the displacement coordinate $s$.
Furthermore, formula~\ref{eq:TOFFBP} can be redefined directly in image space as asymmetrical three-component kernel: one along LOR, which represents TOF function $h(t)$, second -- along $Z$-axis, reflecting the uncertainty of hit position, and the third -- as image-domain ramp filter, which operates on transverse plane along the direction, orthogonal to LOR. In this work, we shall use Gaussian definition for TOF and $Z$ components, as it proved to be consistent with FBP3DRP and was successfully tested on both multi-layer and 2-m-long scanners \cite{shopa2020estimation, moskal2020performance}.

\begin{figure}[h]
\centering
\includegraphics[width=0.49\textwidth]{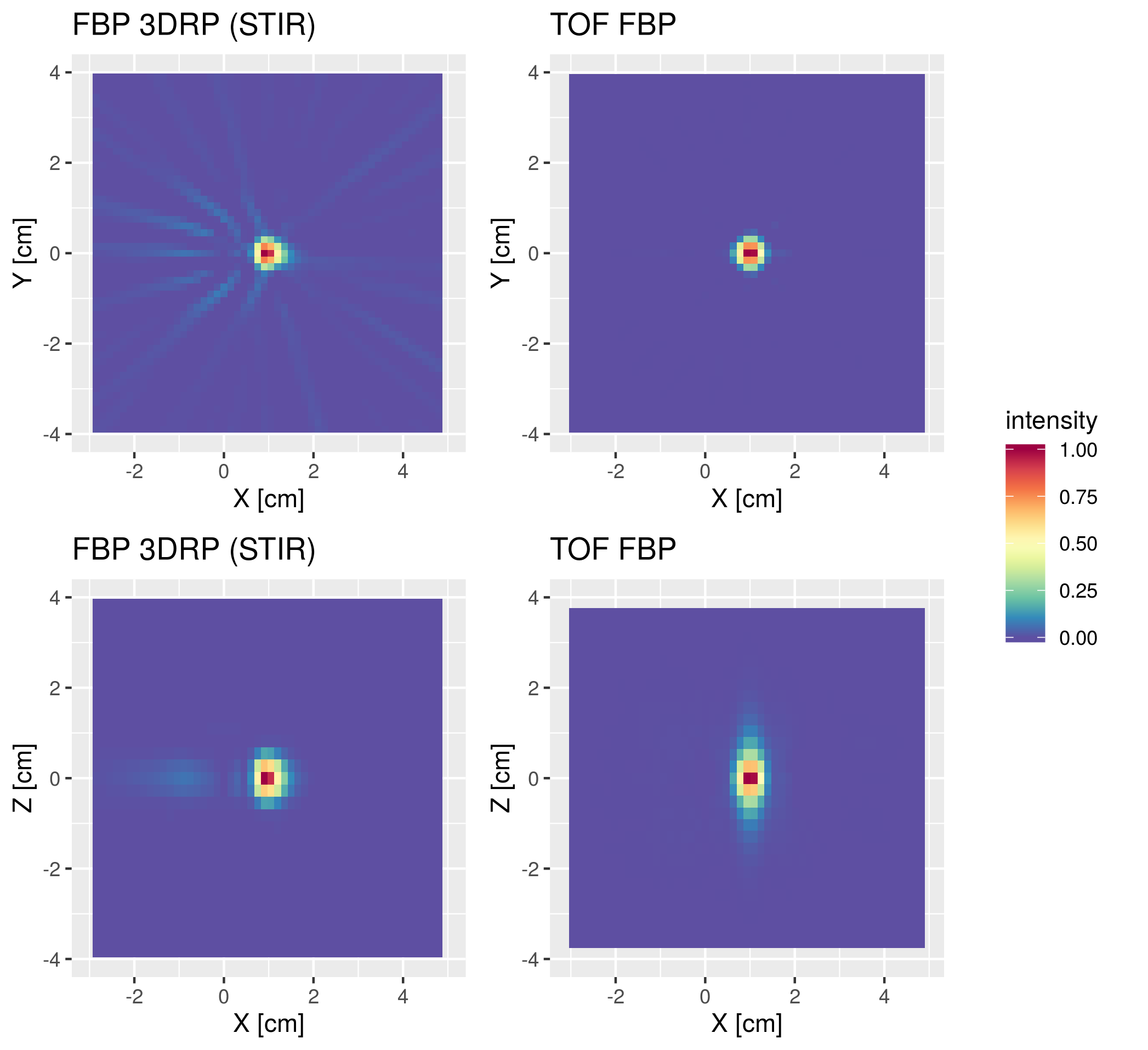}
\includegraphics[width=0.49\textwidth]{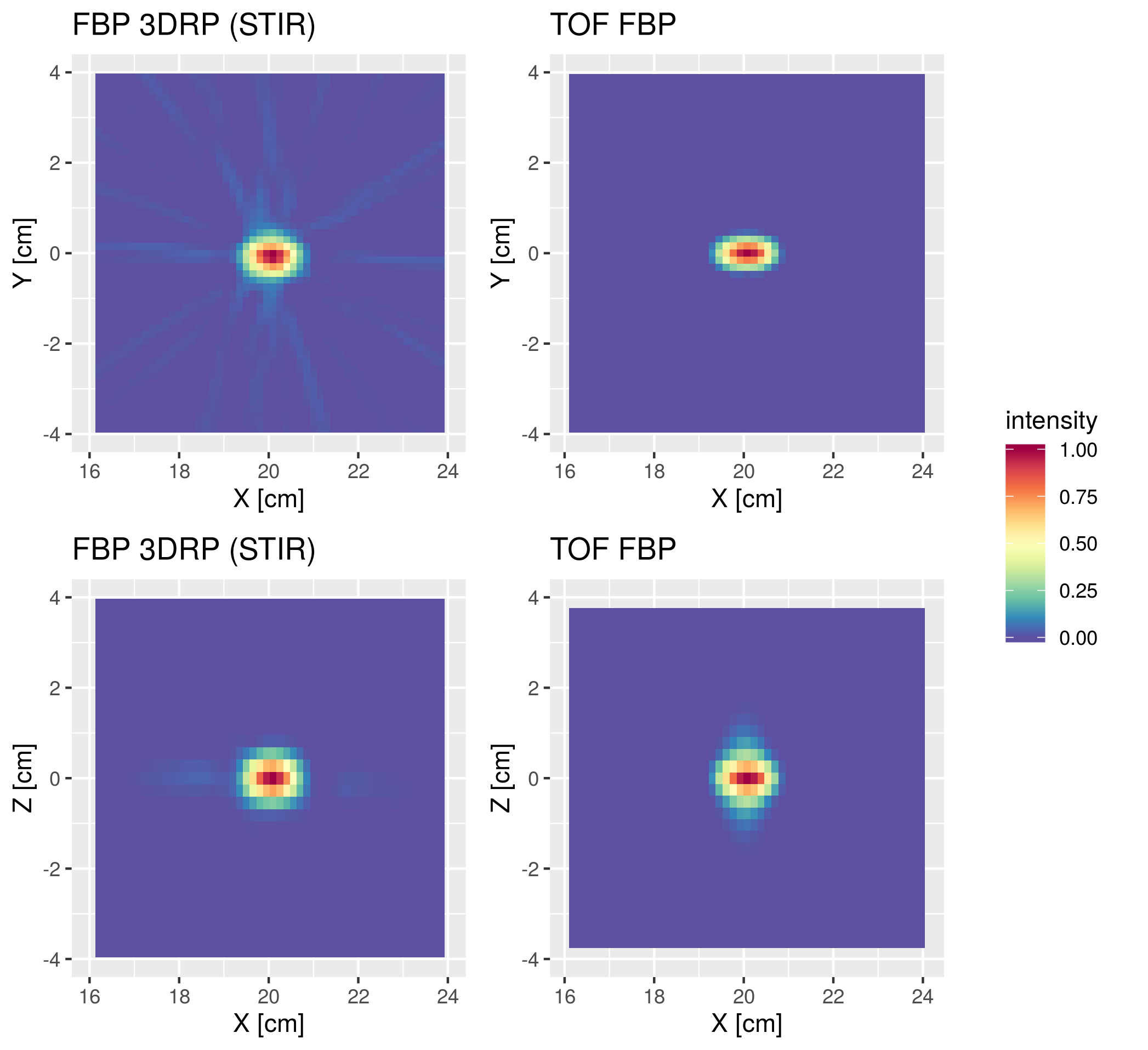}
\caption{
Images of the point source placed in position of (1, 0, 0) cm (2 left columns) and in position of (20, 0, 0) cm (2 right columns) reconstructed with two methods: FBP3DRP and TOF-FBP.
Results for AFOV of 140~cm are shown.
}
\label{fig::sr_methods}
\end{figure}

Each resulting image was additionally corrected by the sensitivity map of the TB-J-PET, generated beforehand in GATE using Monte Carlo method.
For comparison, selected data for 140-cm-long scanner were reconstructed using FBP3DRP.
Since STIR operates with single layer geometries only, in the latter case hits were projected on ideal single layer cylindrical scanner.
%of radius $41.6063$~cm composed of 420 strips of the same 0.6~cm $\times$ 3~cm cross-section.
%Such parameters are closest to the real sampling of view angle and displacement of TB-J-PET in transverse plane.
Furthermore, each strip was axially divided into 256 discrete sections -- maximal value allowed for 1-byte representation of axial coordinate, set in SAFIR \cite{BECKER2017} -- a STIR extension which we employed for the conversion of data into suitable Interfile format.
Each voxel of the reconstructed image for the STIR FBP3DRP had dimensions of 1.50~mm along x and y axes and 2.73~mm along the z axis. % Dxy = 0.149997 cm, Dz = 0.273438 cm
For our TOF FBP implementation a voxel had dimensions of 1.56~mm along x and y axes and 2.59~mm along the z axis.

Comparison of reconstruction results obtained using FBP3DRP and TOF FBP, for 140-cm-long tomograph and the point source placed in position of $(1, 0, 0)$ cm is presented in Fig.~\ref{fig::sr_methods}.

The outcomes are fairly consistent with each other.
Therefore, further on, for the estimation of spatial resolution we will use the TOF FBP which enables image reconstruction with multi-layer geometry without necessity of remapping interaction points onto single layer.

%Though, the typical background artifacts of FBP are more pronounced for STIR in $XY$-plane, because FBP3DRP does not take advantage of TOF information.
%The other algorithm -- TOF FBP -- exhibits much less noise.
%On the other hand, re-projection along with Kinahan and Rogers method \cite{Kinahan1990} favours STIR implementation as it produces slightly better axial resolution.
%It is worth to note, though, that the probability of detector-type scatterings is presumably lower for 420-strip 1-layer tomograph than for 2-layer Digital J-PET (not to mention the factor of WLS).
%Therefore, we may consider that the achieved spatial resolution is the same for both methods within measurement errors.
%As in previous work \cite{moskal2020performance}, axial resolution slightly improves at the distance from the center of the AFOV, while transverse resolution worsens further from the axis of the scanner (\ref{fig::sr_methods}). \\

%\FloatBarrier

\subsection{NEMA-NU2-2018 norm}

\subsubsection{Spatial Resolution}

The standard, approved for the estimation of spatial resolution by National Electrical Manufacturers Association (NEMA) \cite{NEMA:2018}, defines point-like source for the measurements and its positioning inside PET scanner: three transverse locations ($x=1$\,cm, 10\,cm and 20\,cm) and two axial -- at the centre and at $3/8$ of AFOV.
%The activity must be as high as to register at least 100,000 counts.
%The standard also requires filtered back projection (FBP) \cite{Helgason1984} employed as reconstruction algorithm, since it preserves intrinsic resolution properties of the scanner.
%Resolution ought to be reported as FWHM and FWTM measured from orthogonal profiles of the reconstructed source along all three axis.
The spatial resolution (point spread function - PSF) is defined as a~FWHM of the reconstructed image, calculated for each direction separately.
For the image reconstruction the TOF FBP method was utilised as explained above in Sec.~\ref{subsection::TOF-FBP}.

\subsubsection{Scatter Fraction and Noise Equivalent Count Rate}
\label{sub::materials_sf_and_necr}

Scatter Fraction (SF) of the PET scanner quantifies the sensitivity of the detector to scattered radiation.
It is expressed as a~ratio between the rates of scattered coincidences and the sum of scattered and true coincidences: $SF = {S \over {S+T}}$.
It is measured (or simulated) for relatively low source activity, such that the contribution of accidental coincidences is negligible.

NECR is the characteristic that shows the influence of scattered and random coincidences on the performance of the scanner as a~function of the source activity and it is a~measure of the effective sensitivity of the scanner~\cite{Conti2009}.
The NECR is defined as: $NECR = {T^2 \over {T+S+R}}$, where T~stands for the rate of true coincidences, S~--~scattered coincidences, R~--~random (accidental) coincidences.

For both SF and NECR, simulated phantom is a~solid cylinder made of polyethylene with an outside diameter equal to 20.3~cm and length 70~cm.
Parallel to the axis of the cylinder, a~hole with diameter 0.64~cm is drilled in a~radial distance 4.5~cm from the axis of the phantom.
A~line source insert is also made of polyethylene and it is a~tube with the inside diameter 0.32~cm and outside diameter 0.48~cm.
The tube may represent known activity and be placed inside the hole of the phantom.

In presented studies, two methods of estimating SF and NECR were used.
The first one is based on the "true" Monte Carlo information about the photons' propagation that is saved in the GATE output file.
Having this information it is possible to judge if a~coincidence is true, scattered or accidental.

On the other hand, in real measurements not all information about the photons' propagation is available.
Because of that, we apply also a second method as proposed in the NEMA norm, which is based on the analysis of sinograms.

\subsubsection{Sensitivity}

The sensitivity of a~positron emission tomograph is expressed as the rate of true coincidence events T~normalized to the total activity A~of the source.
In order to calculate sensitivity, a~linear 1~MBq source of back-to-back gamma photons with length of 70~cm was simulated along the axis of the scanner in the centre of the AFOV.

\subsubsection{Image Quality}

In order to estimate the image quality, simulation of the IEC phantom with cold and hot spheres was performed with the 140-cm-long AFOV detector. The phantom was positioned in the center of the AFOV. Image quality was calculated based on regions of interest (ROIs) located in two cold and four hot spheres described in NEMA norms \cite{NEMA:2018}. Hot spheres of 10 mm (Sphere 10), 13 mm (Sphere 13), 17 mm (Sphere 17) and 22 mm (Sphere 22) diameter, and cold spheres of 28 mm and 37 mm diameter were simulated. The hot spheres were filled with the activity concentration of 4 times higher than the background region. All spheres centres were positioned in the same transaxial plane located 70 mm from the phantom lid. The 180 mm long cylinder of 51 mm diameter was positioned in the center axis of the phantom. Injection of the 53 MBq of (18)F-FDG dissolved in water was simulated. The scan time was set to 500 seconds.

In post-processing, positions of interaction points in scintillators were smeared with the FWHM of 5~mm (along axis of the detecting chamber), while the TOF resolution was set to 135~ps (FWHM). All the simulations were performed with GATE \cite{sarrut2014review,sarrut2021review}.

The field of view (FOV) of the scanner was set to 50x50x130 cm$^3$. The voxel size was 2.5 mm$^3$ isotropic. Only true coincidences were taken into reconstruction and their number was about 219 mln. The images were reconstructed with the LM-TOF-MLEM algorithm with the CASTOR (version 3.1) toolkit \cite{merlin2018castor}. 20 iterations were used. In the reconstruction, sensitivity and attenuation corrections were included.

Two image quality evaluation metrics were calculated: contrast recovery coefficient (CRC) and background variability (BV). In order to calculate aforementioned parameters, circular region of interest (ROI) was defined on each hot sphere. Furthermore, 12 circular ROIs were also defined on the phantom background. Then, they were replicated to four transaxial slices $\pm$ 1 cm and $\pm$ 2 cm resulted in 60 ROIs defined in the phantom background for each sphere in total. The CRC for each hot sphere with diameter $d$ was calculated as follows:

\begin{equation}
\label{eq:crc}
    CRC=\frac{C_{H,d}/C_{B,d}-1}{4-1},
\end{equation}

where $C_{H,d}$ was the mean counts in the hot sphere and $C_{B,d}$ was the mean of the background ROI counts and 4 in delimiter stands for the true activity ratio between hot spheres and phantom background. The BV for each sphere with diameter $d$ was calculated as follows:

\begin{equation}
\label{eq:bv}
    BV=\frac{S_{d}}{C_{B,d}},
\end{equation}

where $S_{d}$ was the standard deviation of the background ROI counts.

\subsection{Extension of NEMA Norm for TB-PET Scanners}

As the NEMA-NU-2-2018 norm (and its earlier versions) was provided for classical PET scanners with AFOV less than 50~cm, the sources and phantoms for SF, NECR and sensitivity were limited to 70 cm.
However, for total-body PET scanners such sources and phantoms may be not sufficient.
Because of that, in discussion section, results of S, SF and NECR obtained for sources and phantoms elongated to 140~cm and 200~cm are presented.

%==========================================================
\section{Results}
\label{sec::results}
%==========================================================

% ./simulations_manager.py -m aggregate -ss nema -gs djpet-total-body --long-sources
% ./simulations_manager.py -m calculate -ss nema -gs djpet-total-body --long-sources
% ./simulations_manager.py -m plot -ss nema -gs djpet-total-body --long-sources

\subsection{Spatial Resolution}

\begin{figure}[h]
\centering
\includegraphics[width=\textwidth]{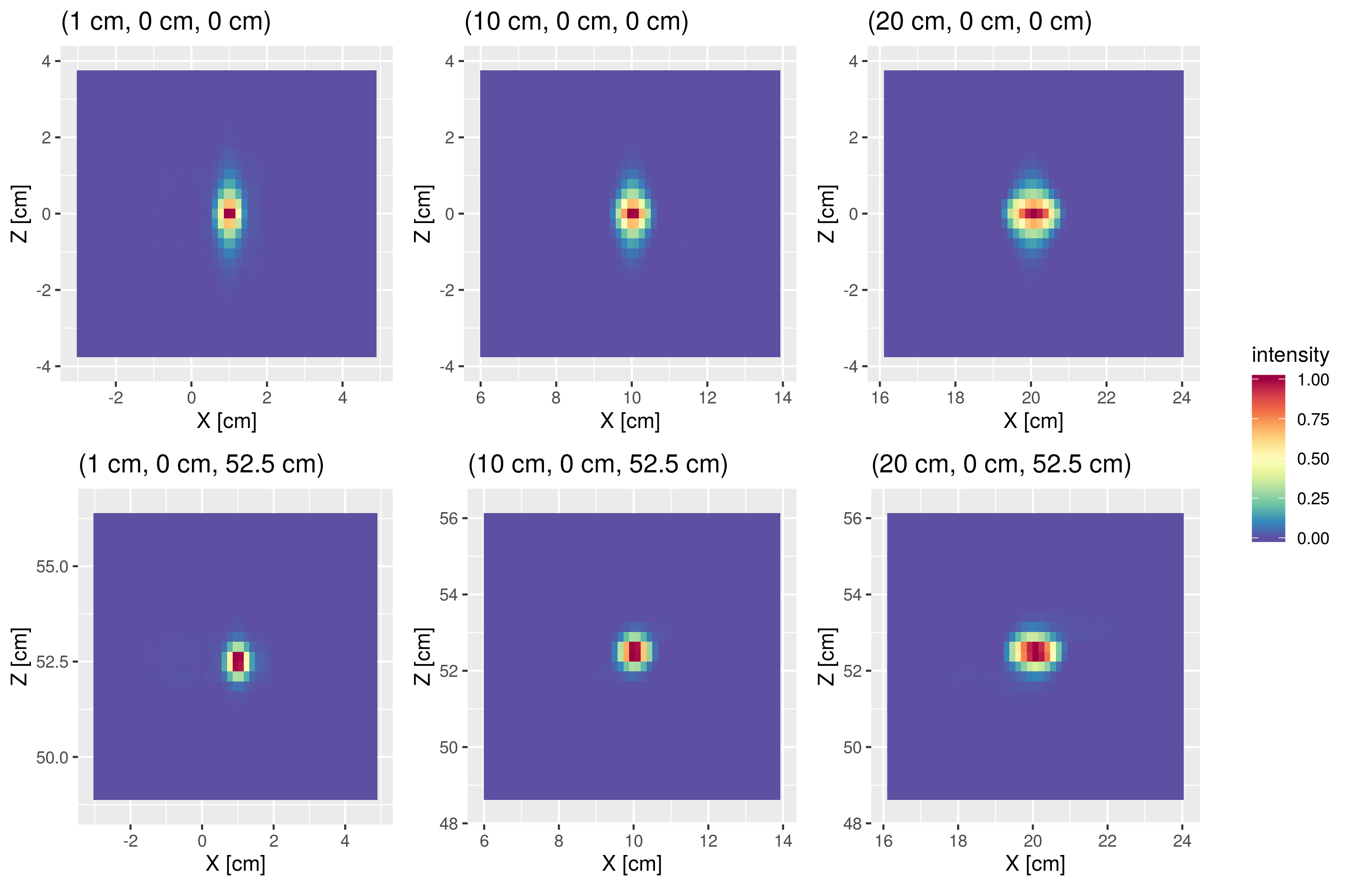}
\caption{
Example of coronal cross section (axial plane) of images reconstructed for point sources positioned as indicated above the pictures (x, y, z).
\label{fig::sr_axial}
}
\end{figure}

\begin{figure}[h]
\centering
\includegraphics[width=\textwidth]{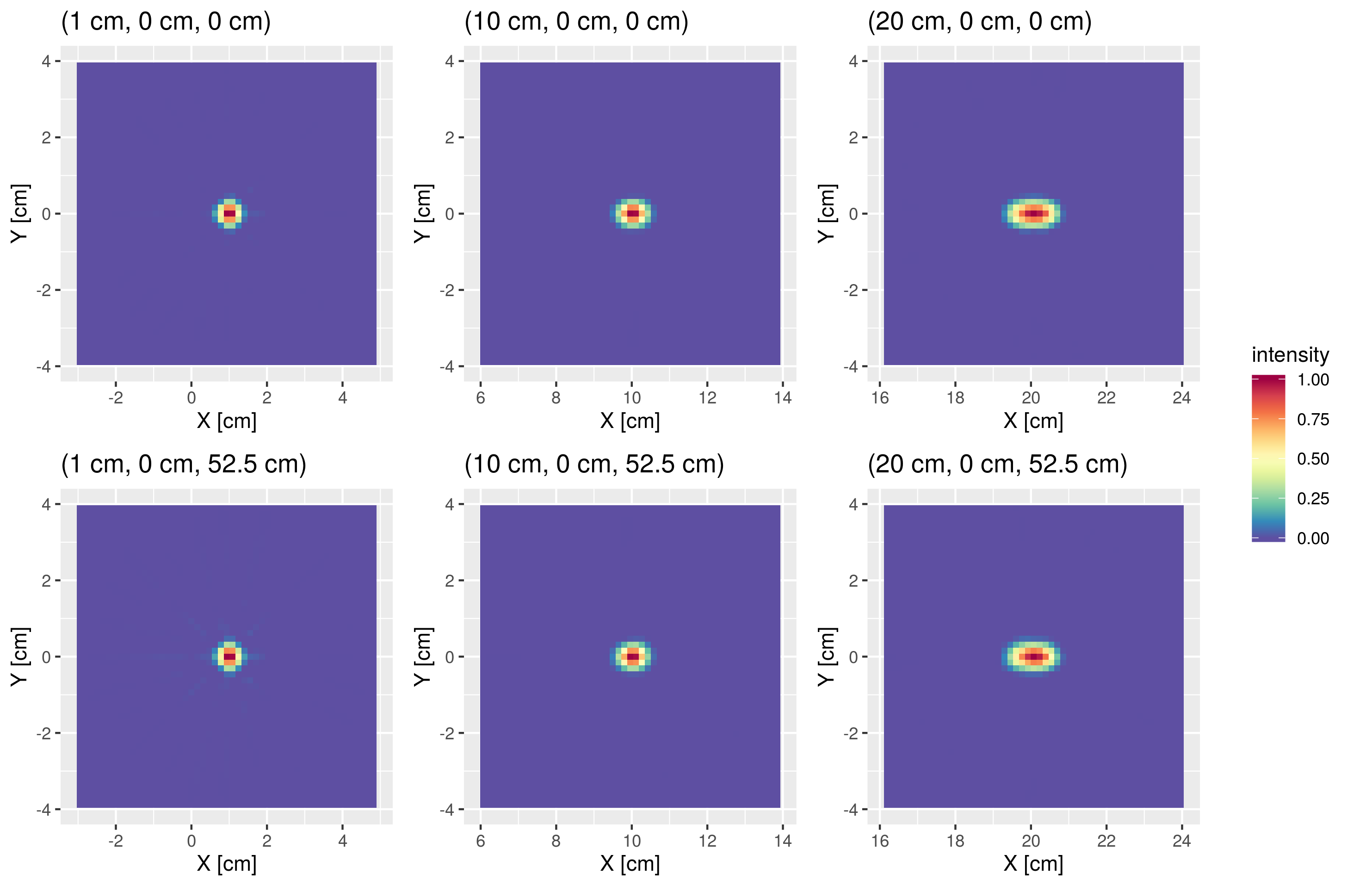}
\caption{
Example of transversal cross sections of images reconstructed for point sources positioned as indicated above the pictures (x, y, z).
\label{fig::sr_transversal}
}
\end{figure}

In order to calculate spatial resolution, TOF FBP reconstruction was utilised as explained in Sec.~\ref{sec::materials}.
Example images of point-like sources reconstructed with TOF FBP for the J-PET scanner with AFOV~=~140~cm are presented in Figs \ref{fig::sr_axial} and \ref{fig::sr_transversal}.
It is visible that achievable spatial resolution is in the order of 5~mm.
Axial spatial resolution (Fig.~\ref{fig::sr_axial}) improves with the axial distance from the center of the scanner, while the spatial radial resolution worsens with the distance from the scanner axis (Fig.~\ref{fig::sr_transversal}).
The values of FWHM and FWTM are given in Tabs~\ref{tab::PSF_ideal_DOI}, \ref{tab::PSF} and \ref{tab::PSF_DOI_1cm}.

\begin{table}
\centering
\includegraphics[width=\textwidth]{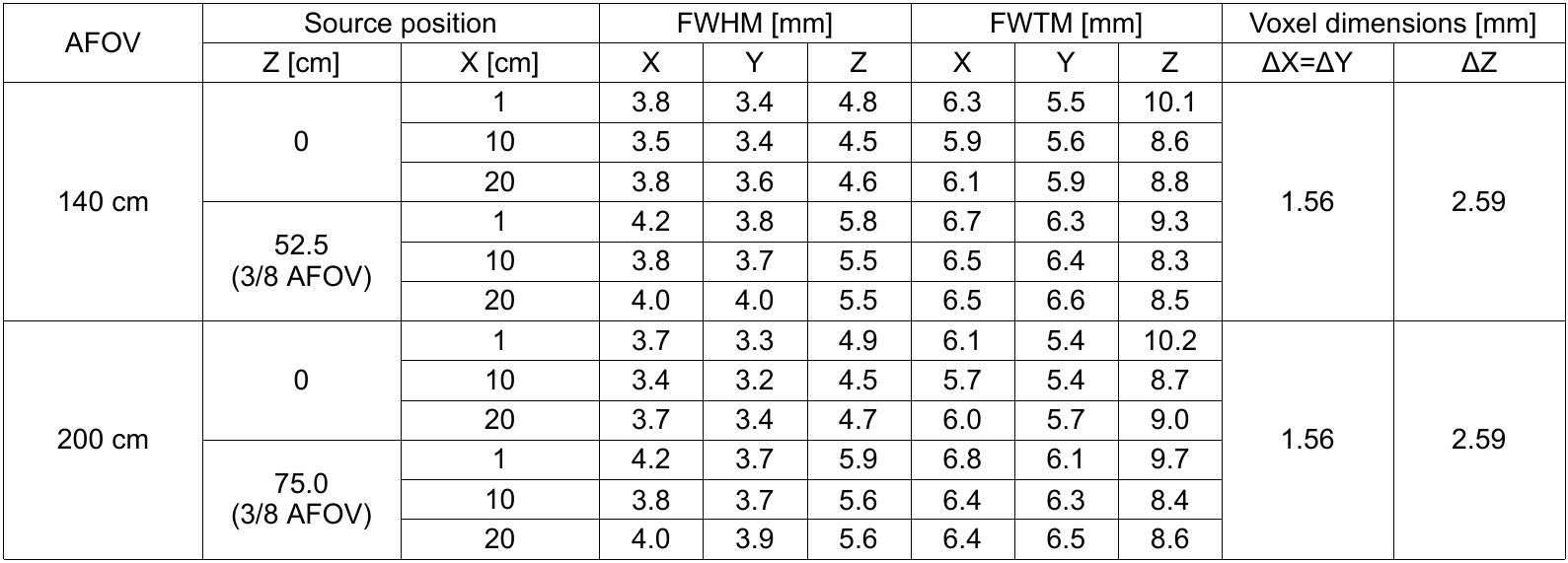}
\caption{
Spatial resolution, expressed as FWHM and FWTM of PSF, determined for AFOV of 140 cm and 200 cm assuming an ideal case that DOI is known.
%Sigmas for TOF reduced by factor 1.5 (0.8-1.0cm): simulated DJPET (24 modules x16 strips in each x2 layers r=40.808027535 cm, 44.308027535 cm)
\label{tab::PSF_ideal_DOI}
}
\end{table}

\begin{table}
\centering
\includegraphics[width=\textwidth]{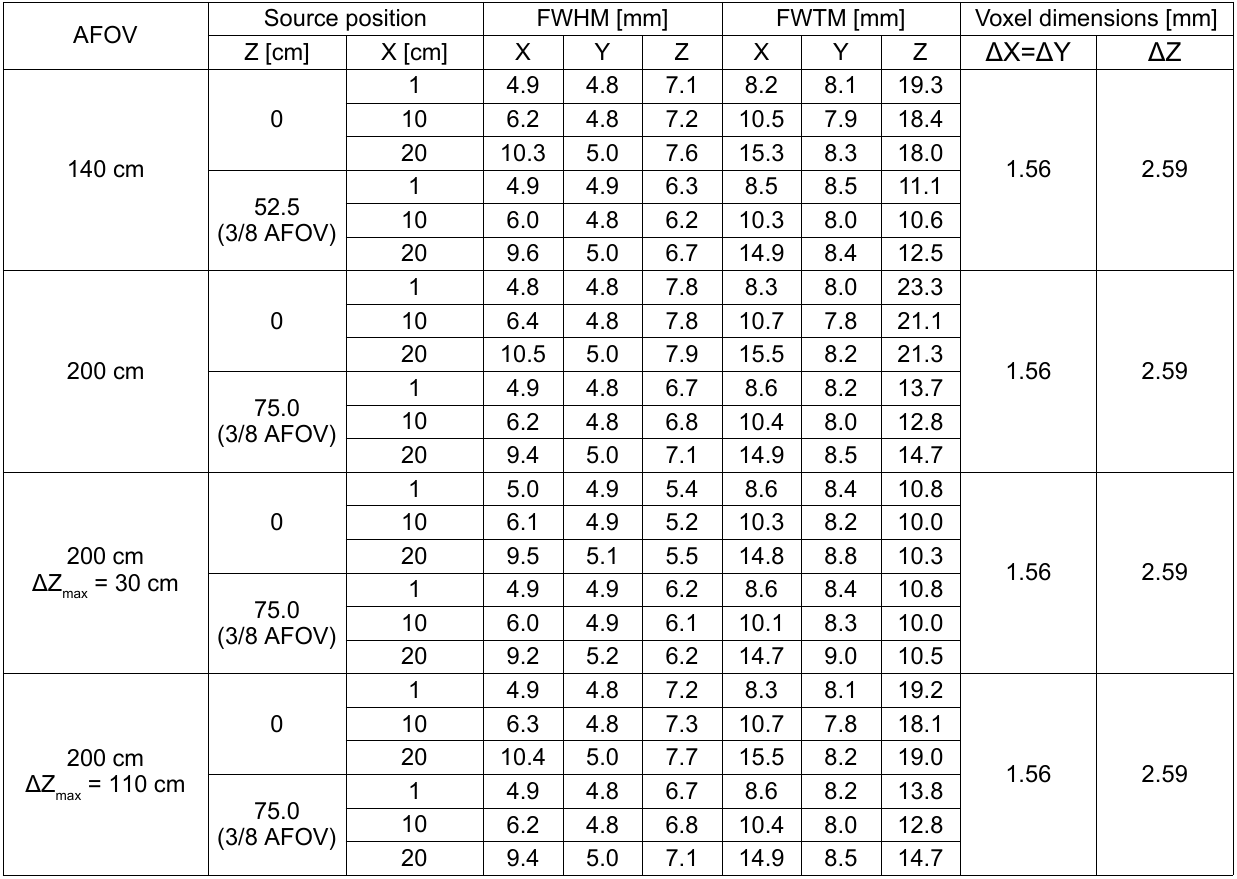}
\caption{
Spatial resolution, expressed as FWHM and FWTM of PSF, determined for AFOV of 140~cm and 200~cm assuming that DOI is not known.
For the case of AFOV~=~200~cm results for the reconstruction with the limited axial distance between hits $\Delta Z_{max}$ are also presented. 
%Sigmas for TOF reduced by factor 1.5 (0.8-1.0cm): simulated DJPET (24 modules x16 strips in each x2 layers r=40.808027535 cm, 44.308027535 cm)
\label{tab::PSF}
}
\end{table}

\begin{table}
\centering
\includegraphics[width=\textwidth]{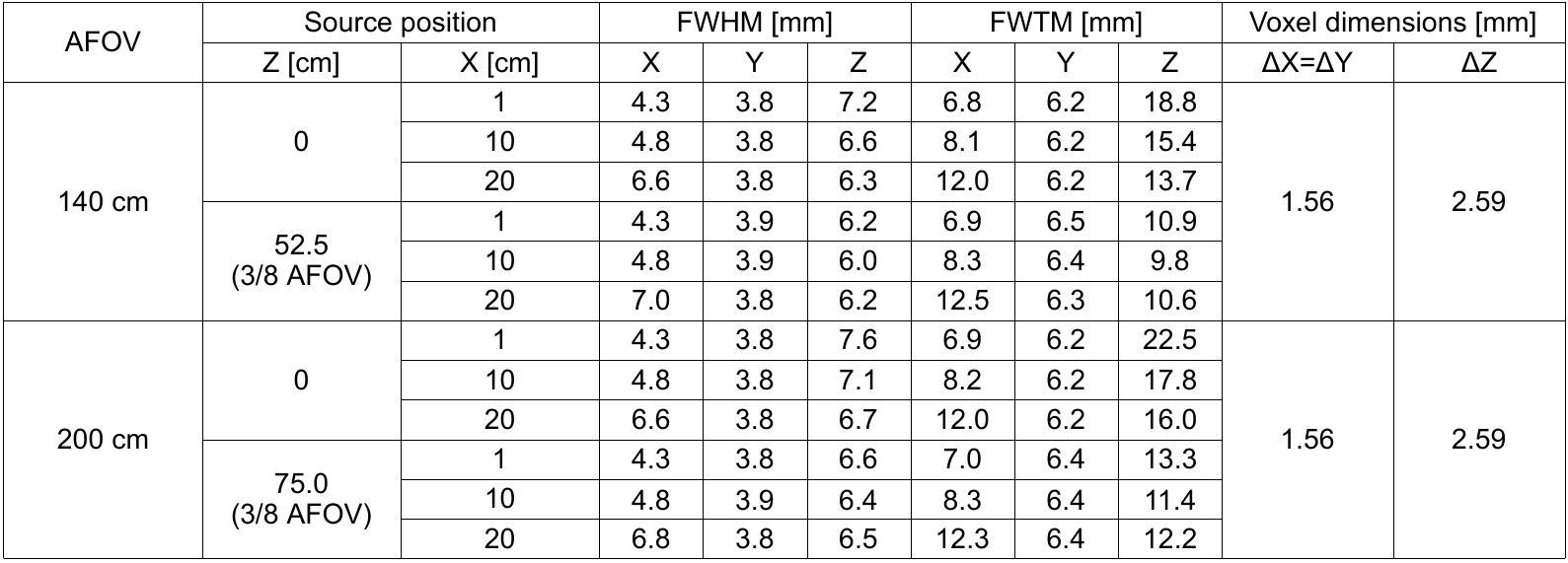}
\caption{
Spatial resolution, expressed as FWHM and FWTM of PSF, determined for AFOV of 140 cm and 200 cm assuming that DOI is determined with the resolution of FWHM(DOI)~=~10~mm. 
%, sigmas for TOF reduced by factor 1.5 (0.8-1.0cm): simulated DJPET (24 modules x16 strips in each x2 layers r=40.808027535 cm, 44.308027535 cm)
\label{tab::PSF_DOI_1cm}
}
\end{table}

In Tab.~\ref{tab::PSF_ideal_DOI} the PSF values are shown for the idealized case of known DOI.
The result indicates that in the center of the scanner, even with the relatively large scintillators cross section of 6~mm times 30~mm, FWHM values of PSF are equal to 3.3~mm, 3.7~mm and 4.9~mm for the TB-J-PET scanner, for radial, tangential and axial directions, respectively.

Tab.~\ref{tab::PSF} presents results obtained assuming that DOI is unknown, and the center of the scintillator in the transversal plane is taken as a position of photon interaction.
In this case, in the center of the scanner, the FWHM values of PSF, achievable for the assumed geometry, are equal or better than 5~mm (radial,tangential) for both AFOV~=~140~cm and AFOV~=~200~cm.
In case of axial PSF the FWHM values of about 7.1~mm and 7.8~mm are obtained.

Tab.~\ref{tab::PSF} also shows results of spatial resolution 
estimated limiting axial (oblique) angle of LOR.
Following Ref.~\cite{Zhang2017} the test was performed restricting maximal axial distance $\Delta Z_{max}$ between each pair of hits to 30~cm and 110~cm.
Obtained results indicate that the axial PSF improves from 7.8~mm to 7.2~mm with limiting axial acceptance distance $\Delta Z_{max}$ to 110~cm, and it improves further to 5.4~mm  for $\Delta Z_{max}$~=~30~cm.

Finally the PSF values for the TB-J-PET were determined assuming that DOI can be reconstructed with the resolution described by the Gaussian function with FWHM of 10~mm (Tab.~\ref{tab::PSF_DOI_1cm}).
In principle, the WLS layer can allow for determination of DOI ~\cite{patentWLS}, yet this solution was not yet tested experimentally.
The result presented in Tab.~\ref{tab::PSF_DOI_1cm} shows an improvement of radial, tangential and axial values of PSF with respect to the case when the DOI is unknown.
Specifically the tangential spatial resolution is improved, and it is lower than 4~mm for all positions defined in the NEMA norm. 

%\begin{figure}[h]
%\centering
%\includegraphics[width=0.5\textwidth]{SmearHitByDOI.png}
%\caption{
%Pictorial visualisation of the hit position smearing method on the basis of one of strips (the arrow is directed along the radius of the detecting chamber);
%yellow star shows original position of the hit while the white one the new smeared position;
%$x$ and $y$ coordinates of each hit were smeared along the radius of the detecting chamber.
%}
%\label{fig::smearing}
%\end{figure}

%Intuitively, it seems that resolutions along x and y should be the same.
%Differences are due to a rectangular shape of a single strip (30 mm x 6 mm) and due to modular construction of the investigated Digital J-PET detector.

\FloatBarrier

%==========================================================

\subsection{Scatter Fraction and Noise Equivalent Count Rate}

\begin{figure}[h]
\centering
\includegraphics[width=0.49\textwidth]{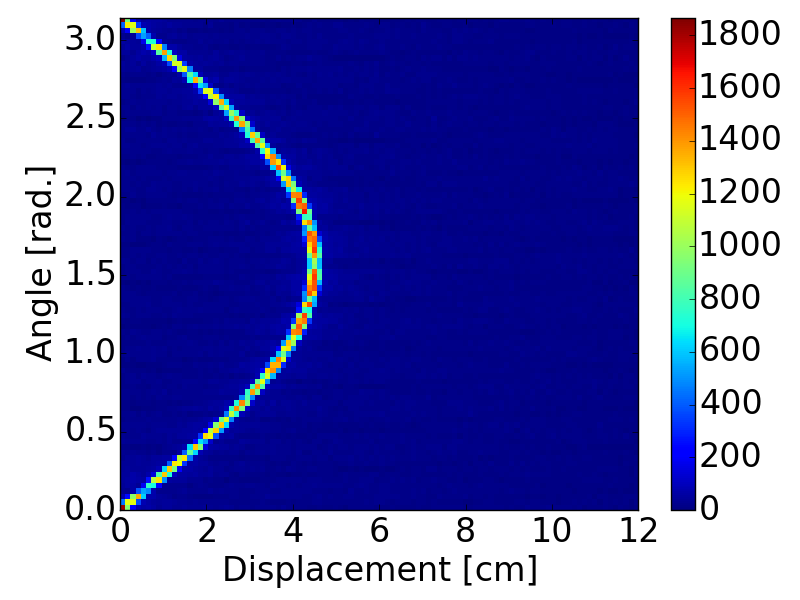}
\includegraphics[width=0.49\textwidth]{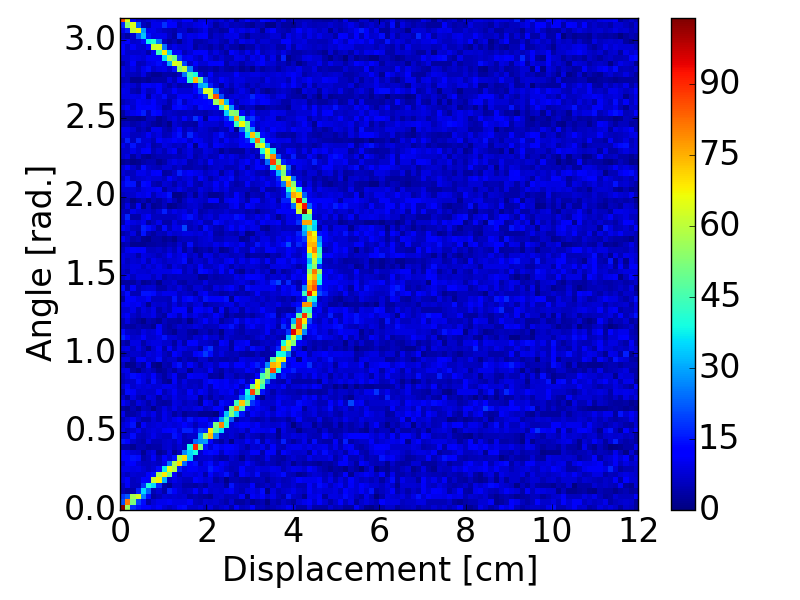}
\includegraphics[width=0.49\textwidth]{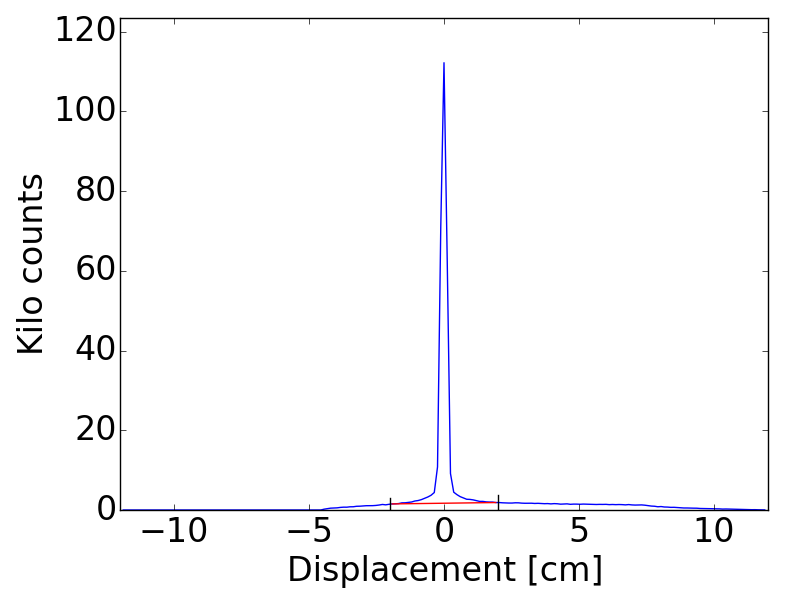}
\includegraphics[width=0.49\textwidth]{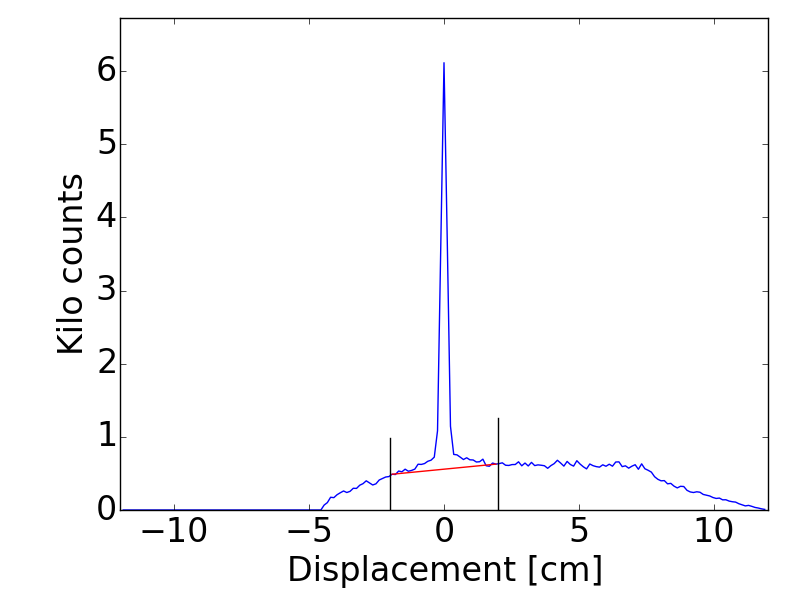}
\caption{
Example NECR simulations result for 200-cm-long prototype with the 70-cm-long source;
(left) 1 MBq total activity, (right) 2000 MBq total activity;
(top) sinograms, (bottom) aligned and summed sinograms, used to differentiate the true and the false (both scattered and accidental) coincidences.
Coincidences over the red line are treated as true coincidences, while those below as scattered and accidental ones.
\label{fig::sinograms}
}
\end{figure}

As it was described in Sec.~\ref{sub::materials_sf_and_necr} both SF and NECR were estimated using two different methods.
In the case of method based on sinograms, the detecting chamber is divided into 1 cm slices and a~sinogram is calculated for each slice.
After alignment and processing (see Fig.~\ref{fig::sinograms}) the number of true coincidences (T) was disentangled from scattered and random (S+R) according to the prescription given in the NEMA norm (see lower panel of Fig.~\ref{fig::sinograms}).

\begin{figure}[h]
\centering
\includegraphics[width=0.49\textwidth]{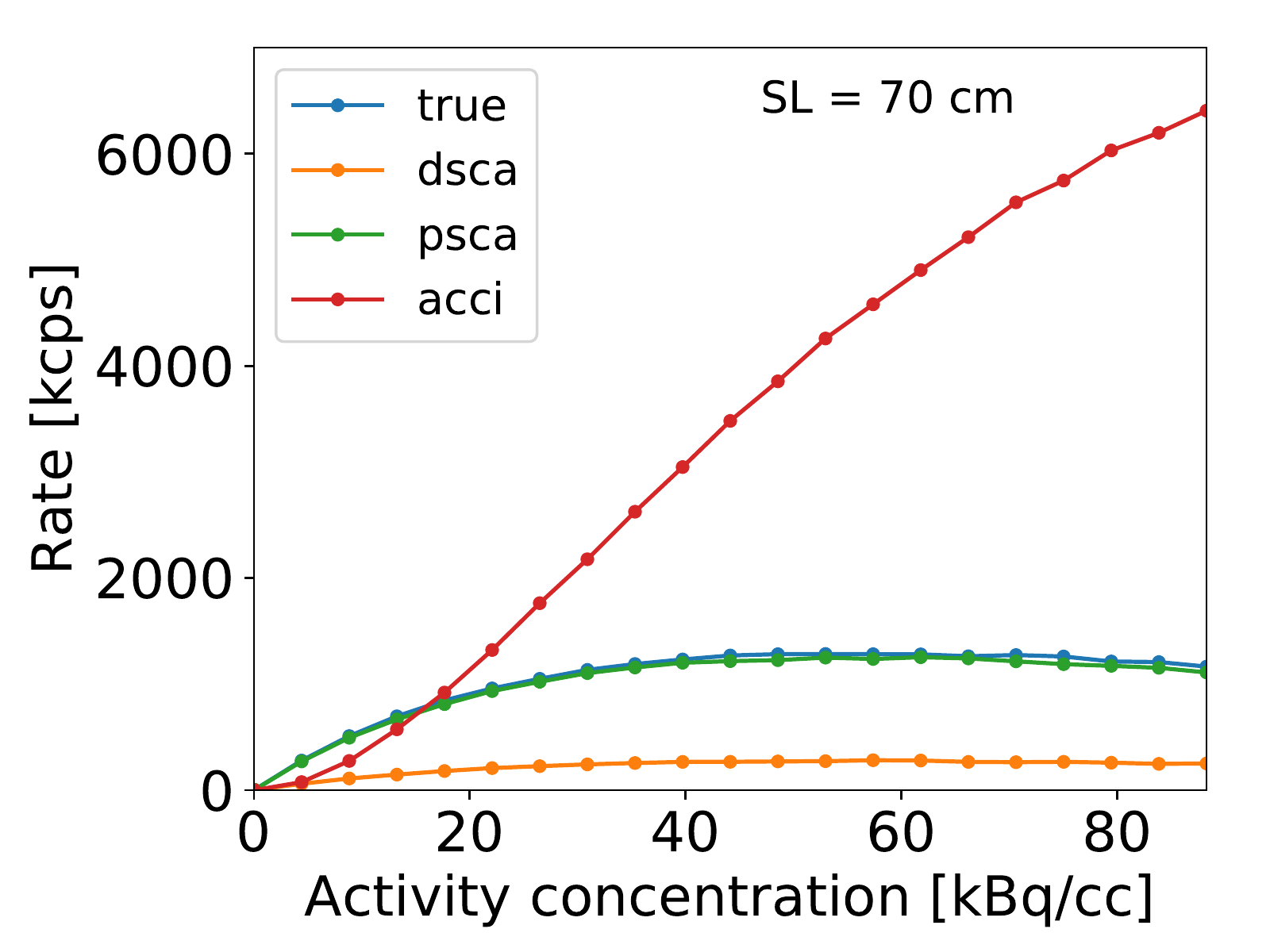}
\includegraphics[width=0.49\textwidth]{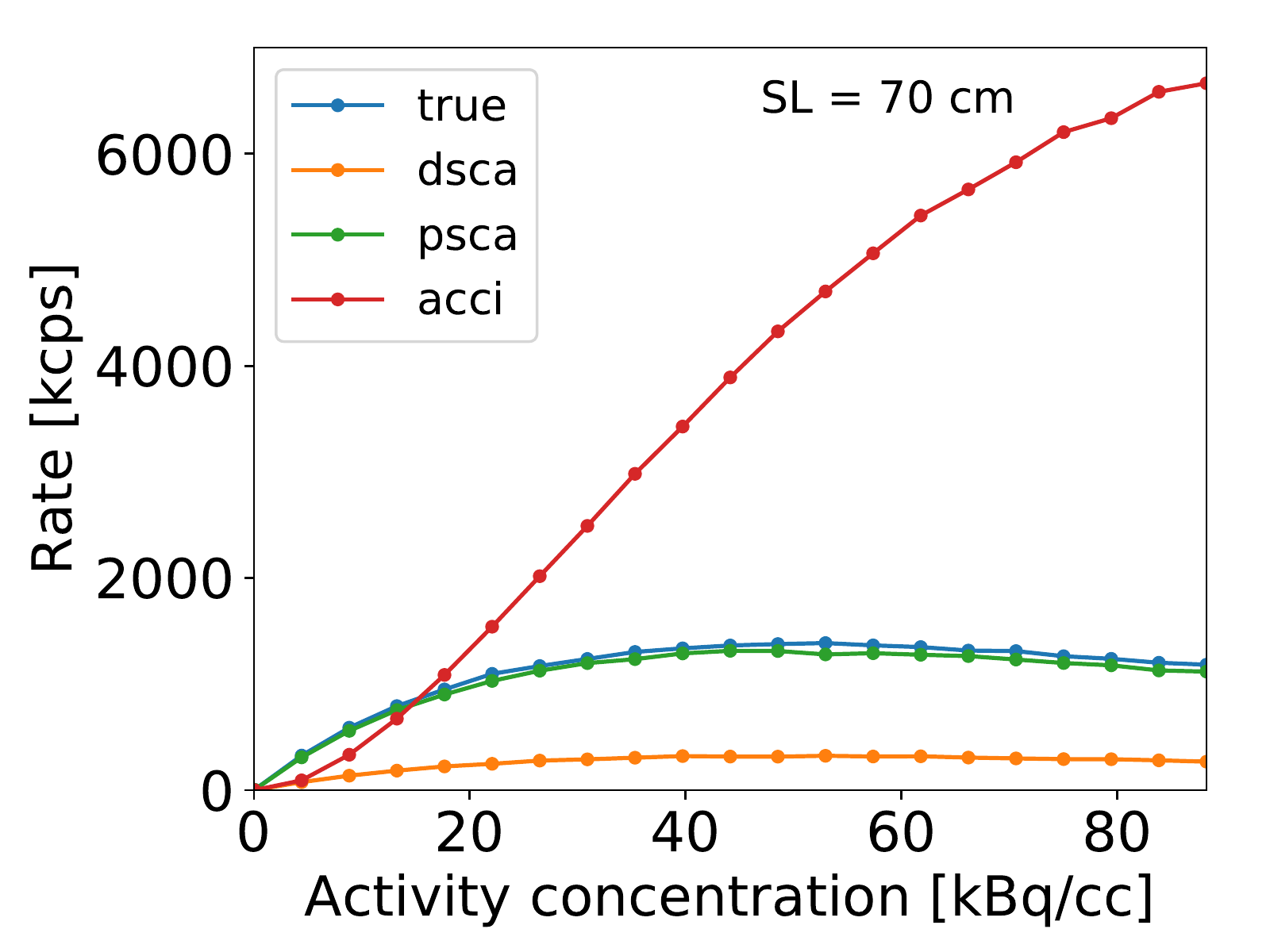}
\includegraphics[width=0.49\textwidth]{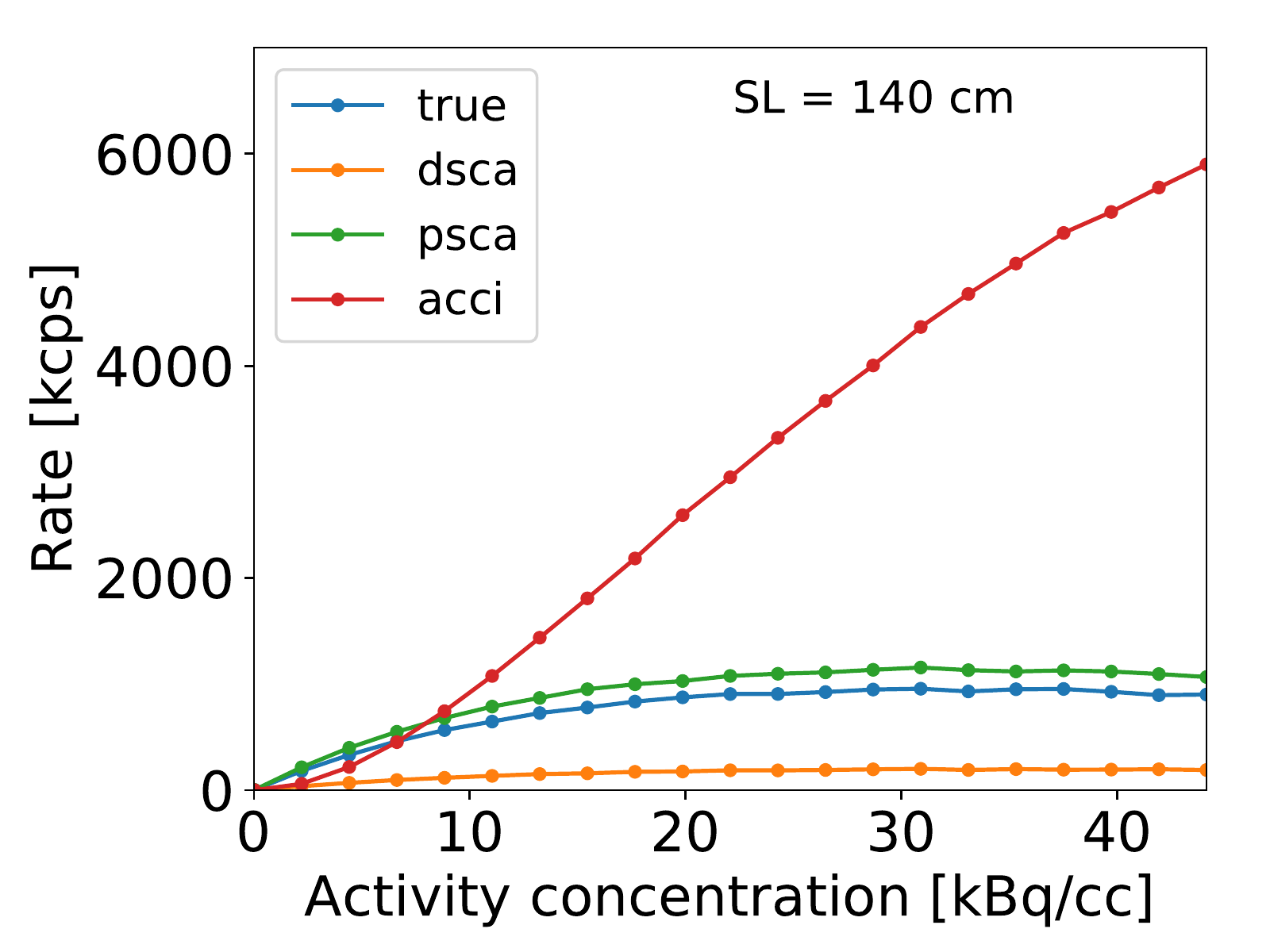}
\includegraphics[width=0.49\textwidth]{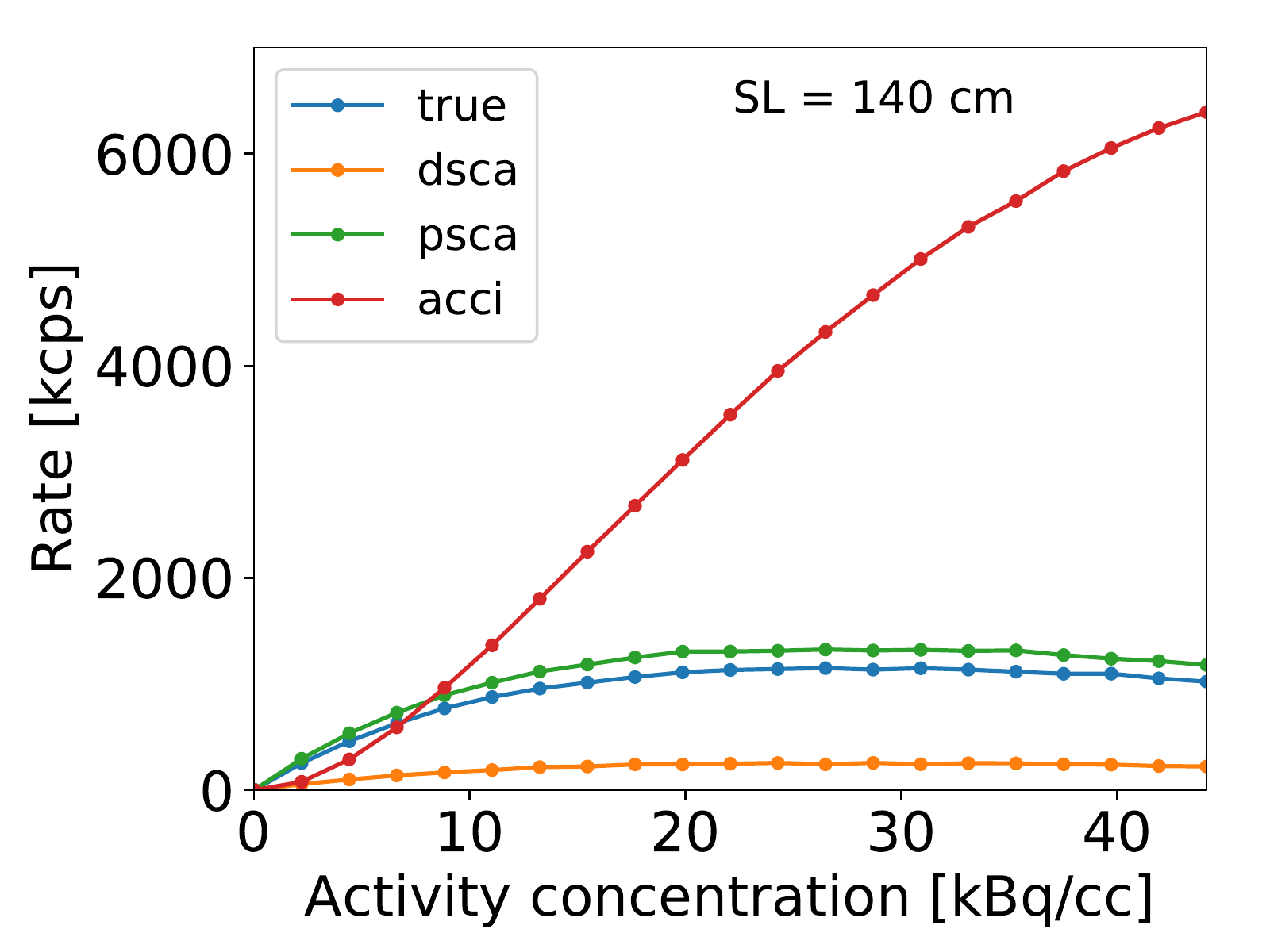}
\includegraphics[width=0.49\textwidth]{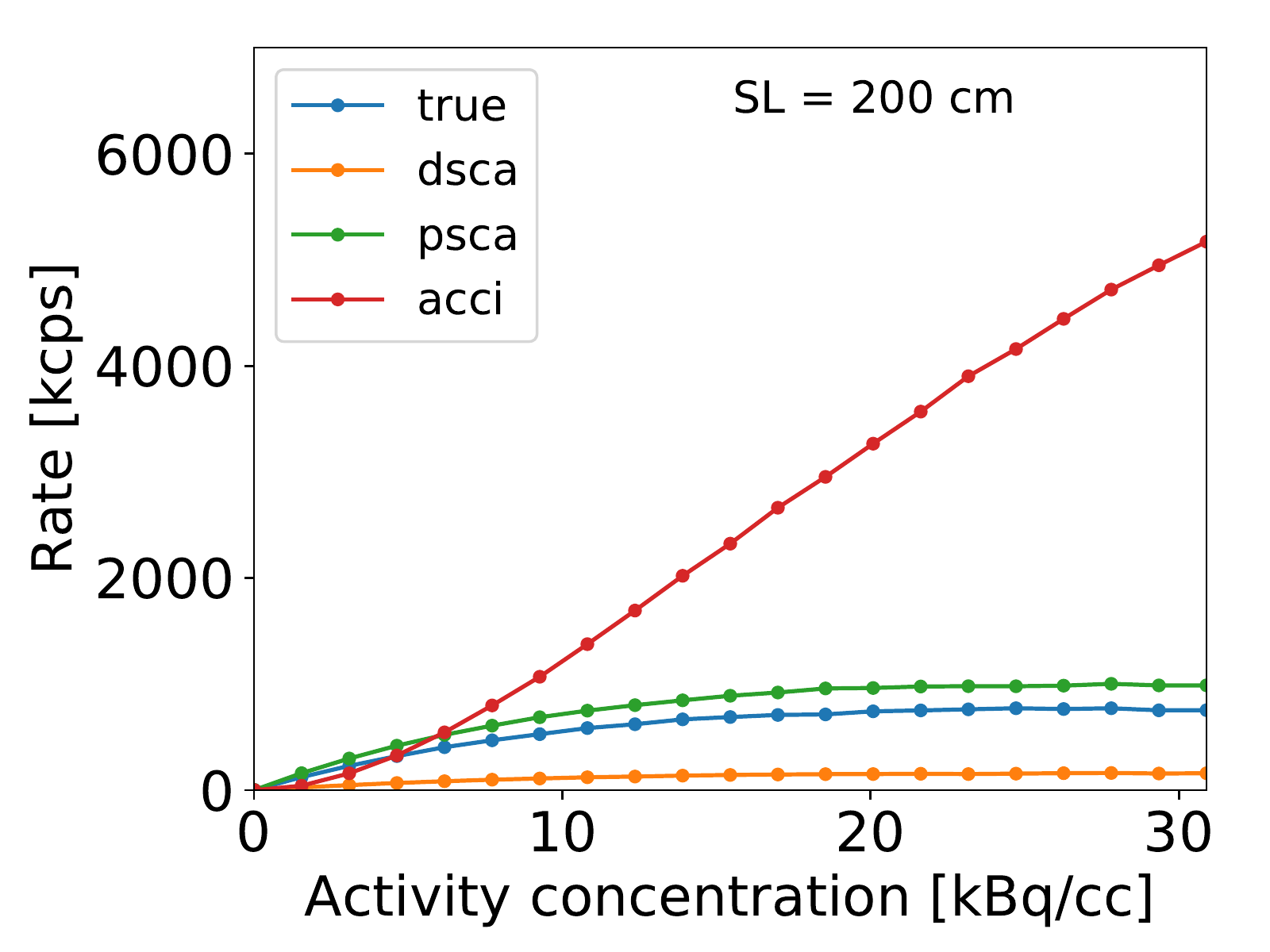}
\includegraphics[width=0.49\textwidth]{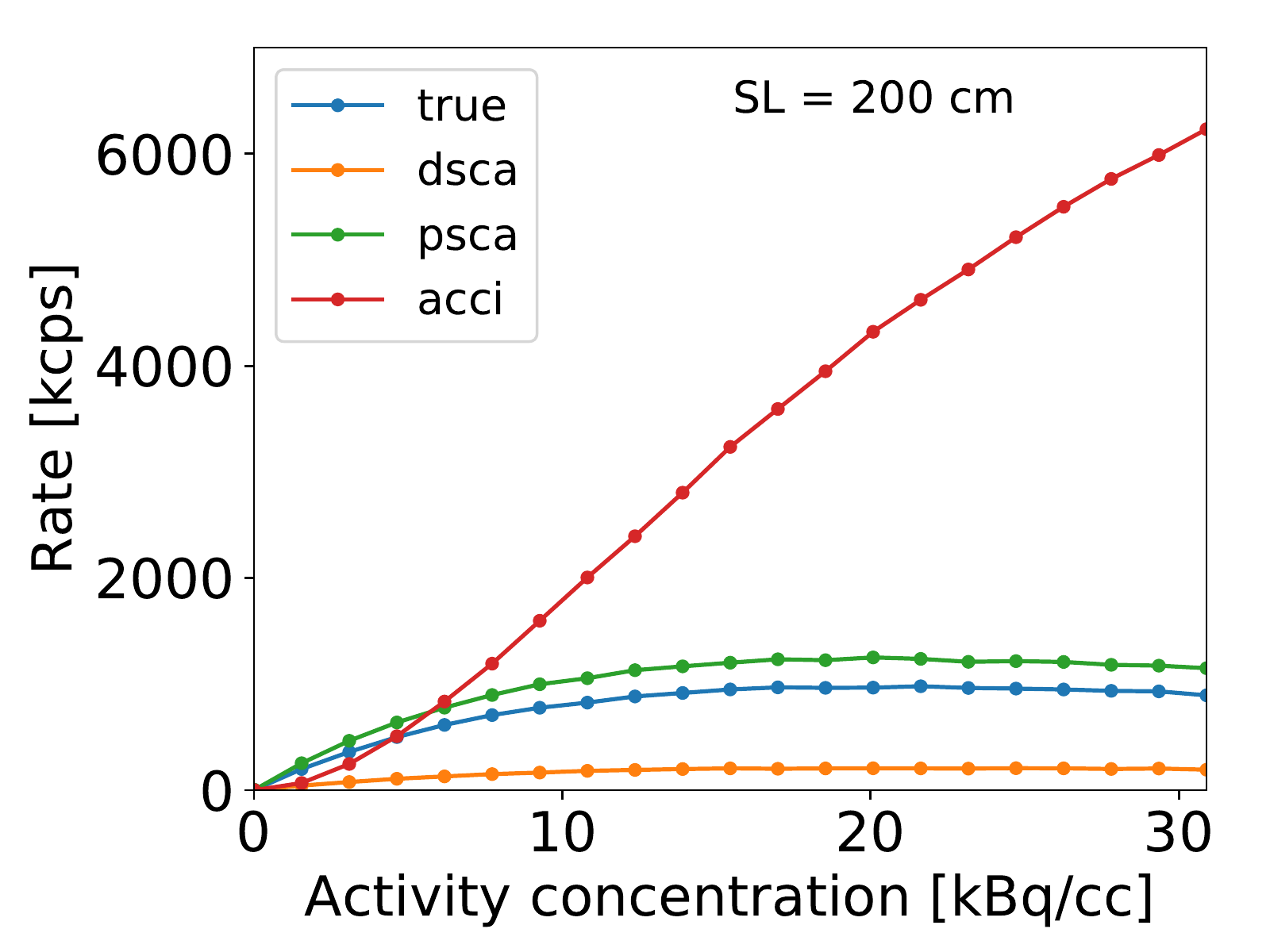}
\caption{
Rates of coincidences based on true Monte Carlo: (left) for AFOV = 140 cm, (right) for AFOV = 200 cm.
Results for the source and phantom lengths of SL~=~70~cm (upper panel), SL~=~140~cm (middle panel) and SL~=~200~cm (lower panel) are shown. Results for true, detector scatttered (dsca), phantom scattered (psca) and accidental (acci) coincidences are indicated.
\label{fig::rates}
}
\end{figure}

\begin{figure}[h]
\centering
\includegraphics[width=0.49\textwidth]{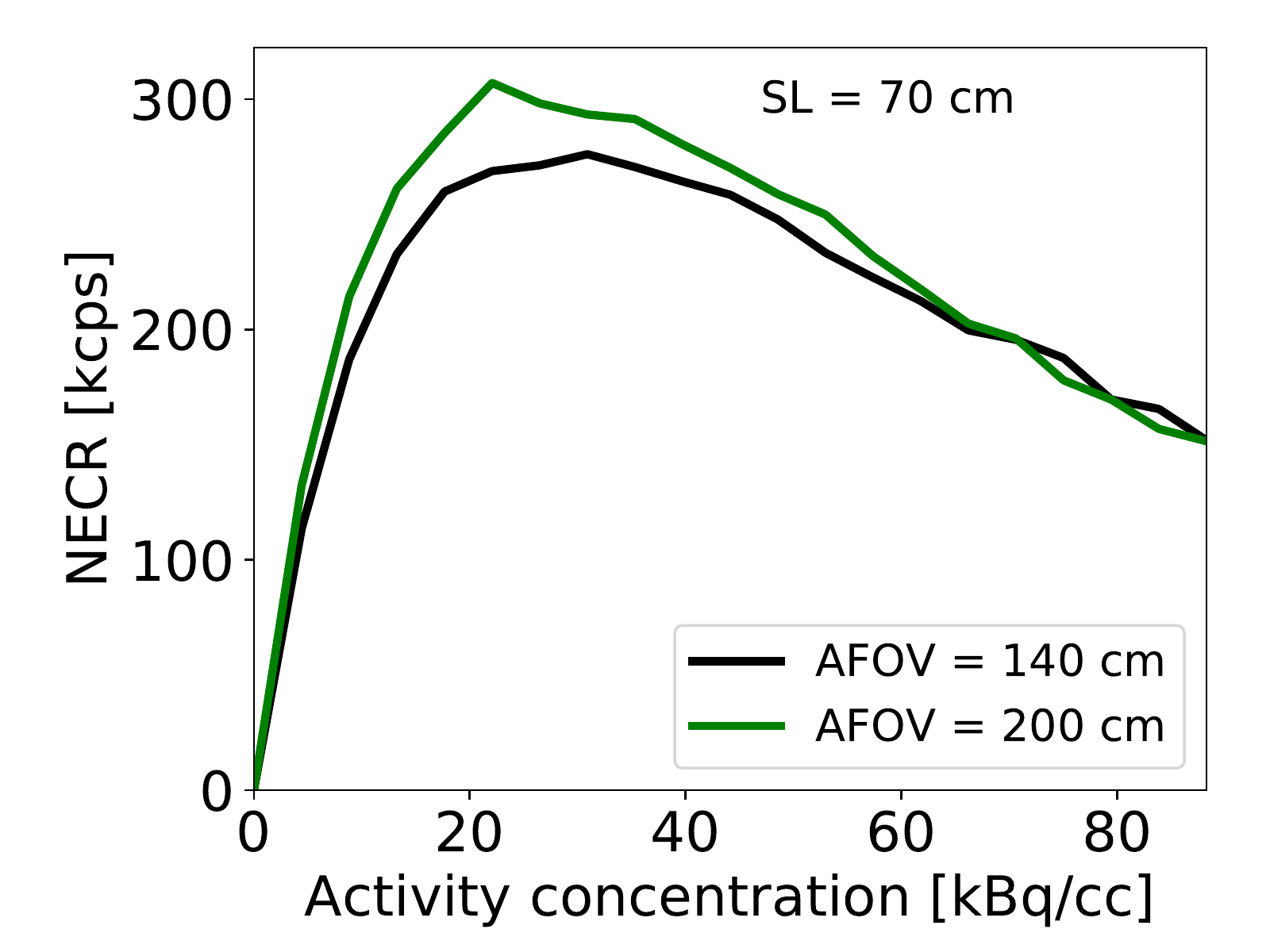}
\includegraphics[width=0.49\textwidth]{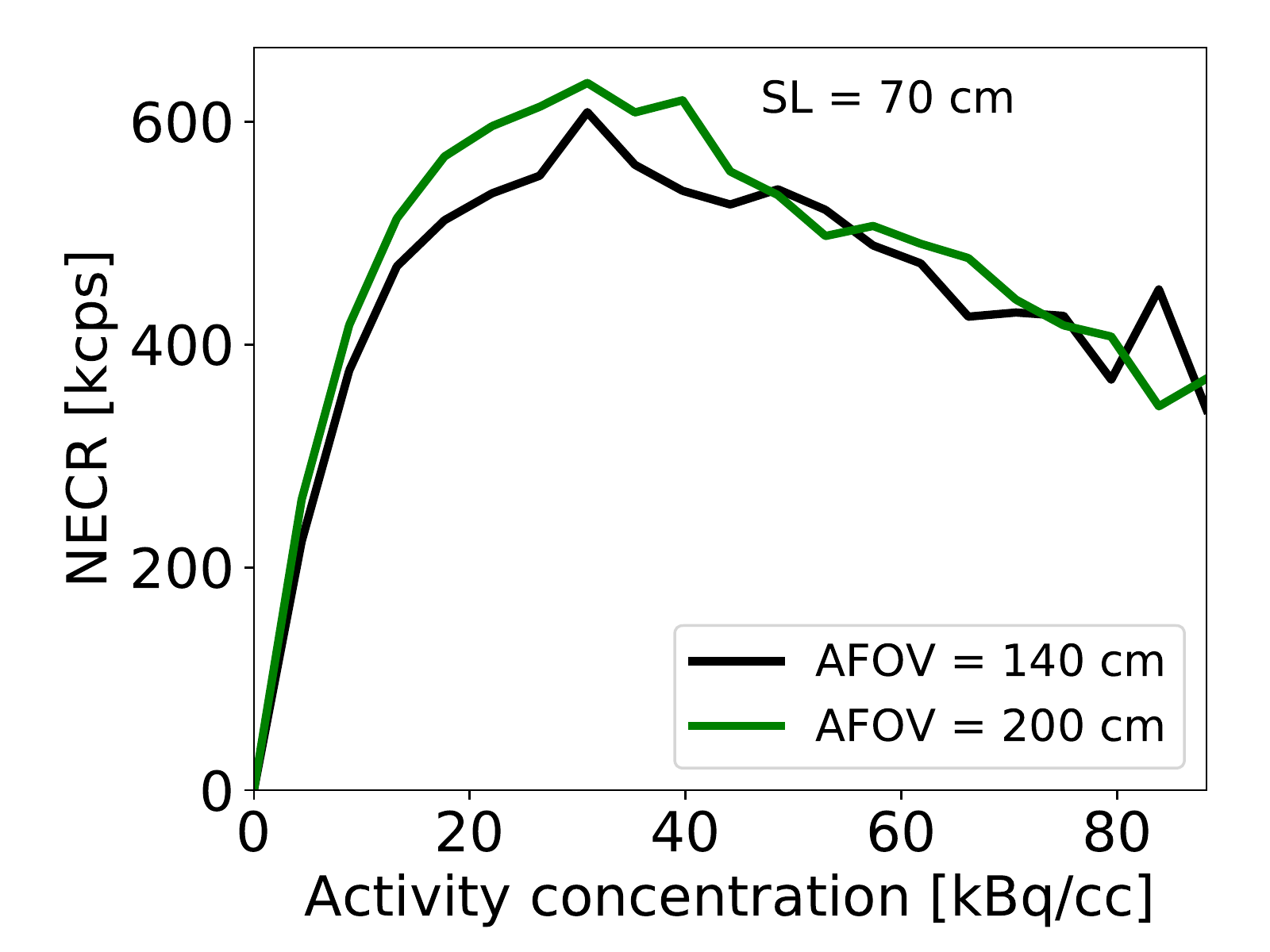}
\includegraphics[width=0.49\textwidth]{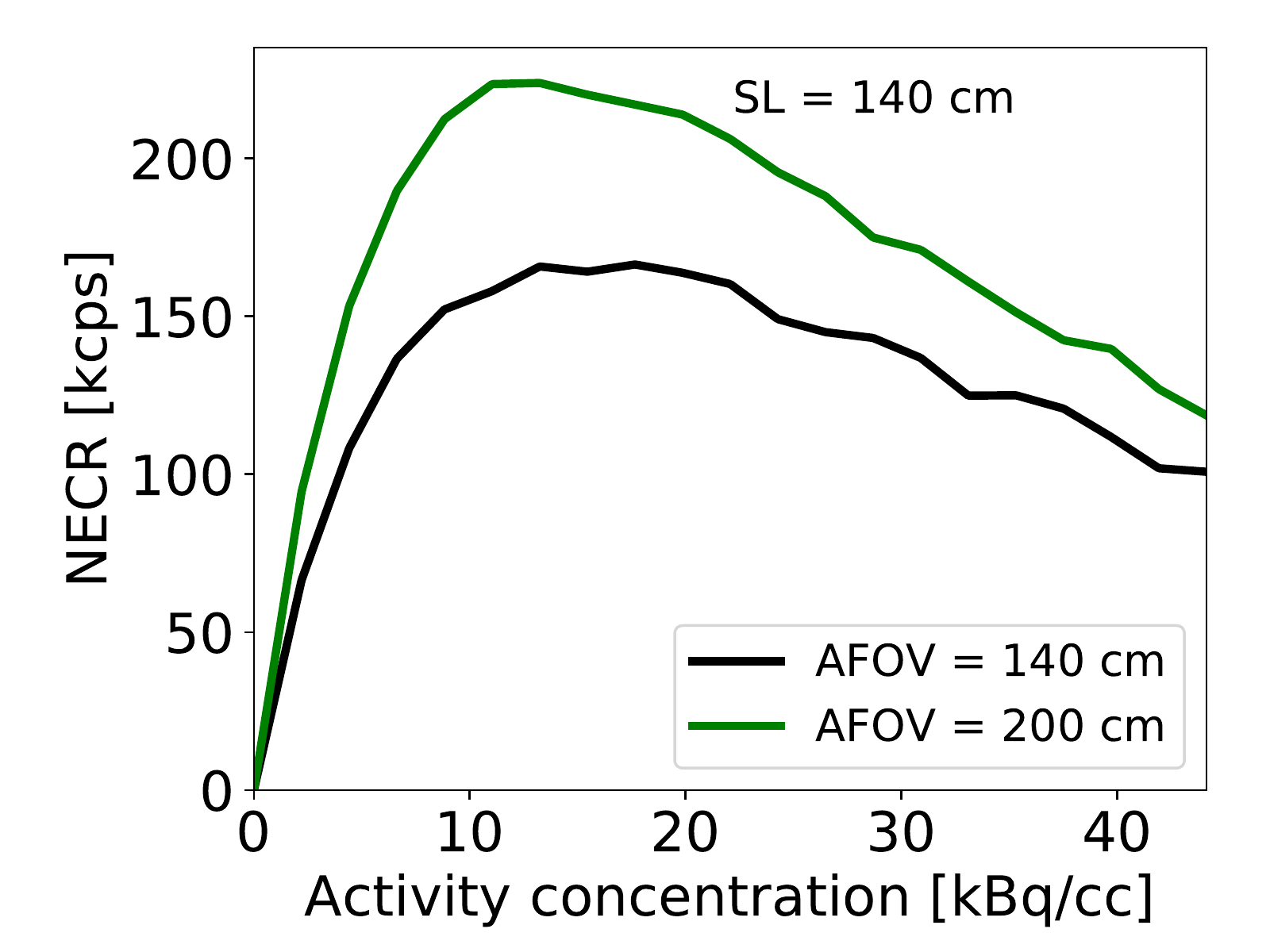}
\includegraphics[width=0.49\textwidth]{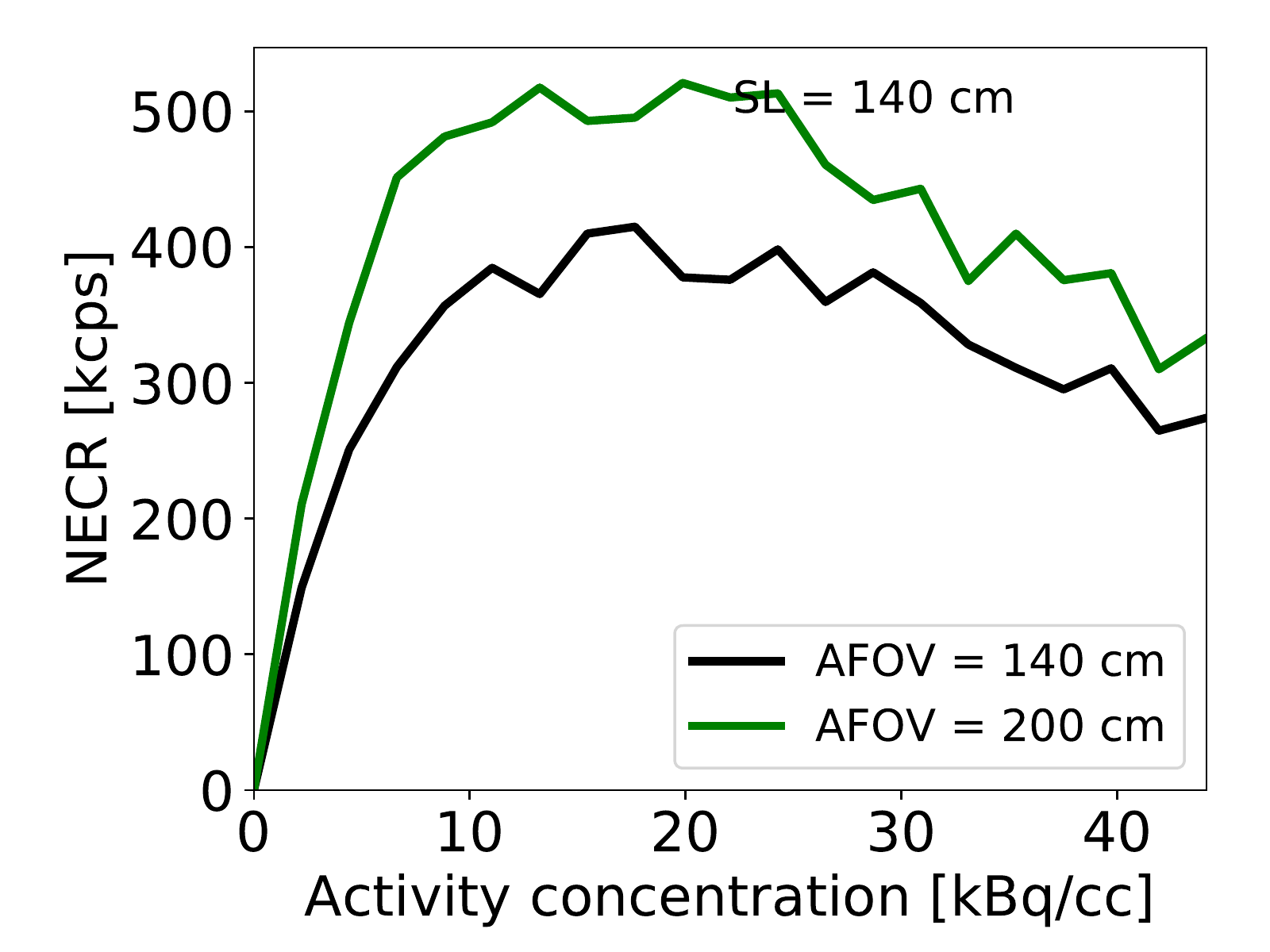}
\includegraphics[width=0.49\textwidth]{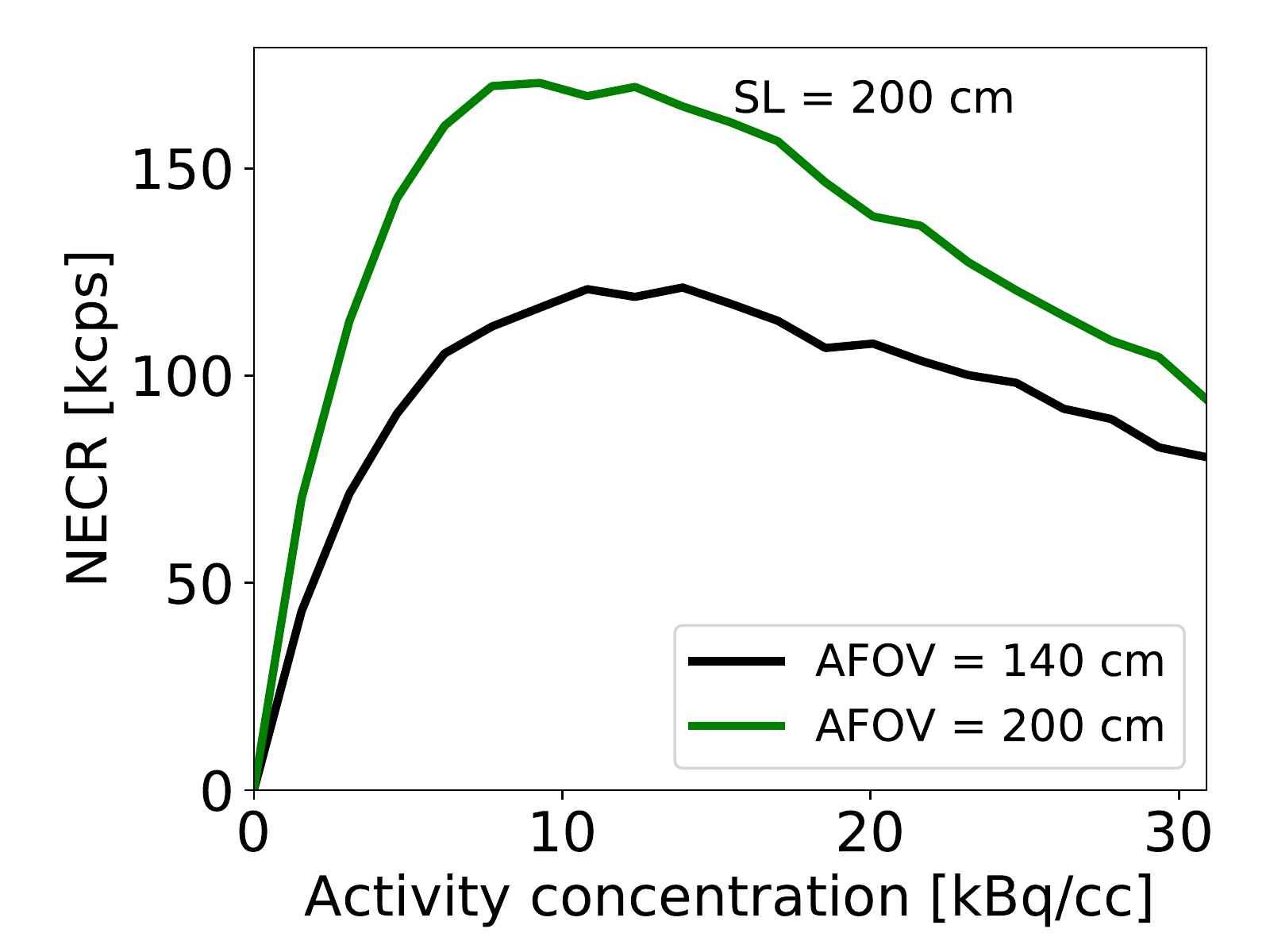}
\includegraphics[width=0.49\textwidth]{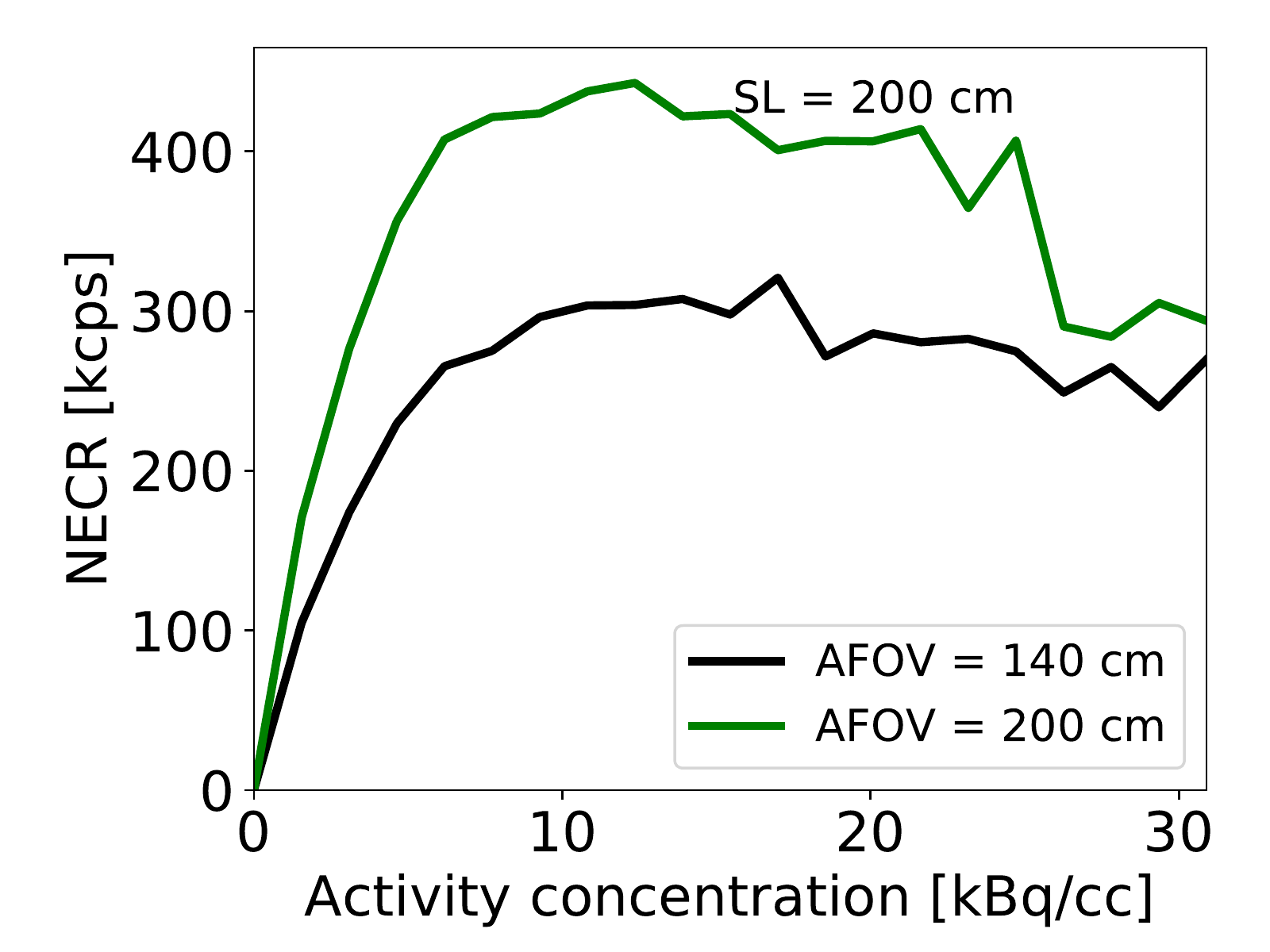}
\caption{
NECR as a function of activity concentration determined (left) based on true Monte Carlo (counters), (right) based on sinograms analysis.
Results for (top) 70~cm long source, (middle) 140~cm long source, (bottom) 200~cm-long source are presented.
\label{fig::necr}
}
\end{figure}

To calculate SF and NECR based on the information available from the Monte Carlo simulations, firstly the rates of subsequent types of coincidences were calculated.
These rates for {\it true}, {\it detector-scattered}, {\it phantom-scattered} and {\it accidental} coincidences are shown in Fig.~\ref{fig::rates} as a function of the activity concentration, calculated as the total source activity divided by the phantom volume.
The determined rates were used to calculate the SF and the NECR as a function of the activity concentration.
Results for SF and NECR are presented in Tab.~\ref{tab::sf_results} and Fig.~\ref{fig::necr}, respectively.
The SF amounts to about 36~\% when using a standard length (recommended by NEMA-NU-2018 norm) of the phantom and source.
There is no significant difference between obtained SF values for two studied lengths of the detector.
Yet, the differences between SF values calculated with two methods are significant.
It is due to the differences in the data selection.
Narrower data cut in case of sinograms limits the number of false coincidences taken into account.
A~similar effect, namely that the SF obtained when using sinograms is smaller than the one obtained using {\it true Monte Carlo} method, was also observed in Ref.~\cite{Yang2015}.

\begin{table}[!htb]
\begin{center}
\footnotesize
\begin{tabular}{|c|c|c|c|}
\hline
	\textbf{Method} &
	\textbf{AFOV = 140 cm} &
	\textbf{AFOV = 200 cm} \\
\hline
	Sinograms, SL = 70 cm &
	35.6\% &
	36.2\% \\
\hline
	True MC, SL = 70 cm &
	54.2\% &
	54.3\% \\
\hline
	Sinograms, SL = 140 cm &
	37.4\% &
	37.6\% \\
\hline
	True MC, SL = 140 cm &
	58.4\% &
	58.2\% \\
\hline
	Sinograms, SL = 200 cm &
	38.2\% &
	38.0\% \\
\hline
	True MC, SL = 200 cm &
	60.0\% &
	59.9\% \\
\hline
\end{tabular}
\caption{
Scatter fraction determined for TB-JPET with AFOV of 140~cm and 200~cm.
Results obtained based on sinograms and on {\it true Monte Carlo} are presented. 
Table includes estimations done for source and phantom length (SL) of 70~cm, 140~cm and 200~cm.
}
\label{tab::sf_results}
\end{center}
\end{table}

The NECR characteristics are presented in Fig.~\ref{fig::necr}.
Like in case of SF, the method based on the analysis of sinograms provides different values for NECR than obtained when using a {\it true Monte Carlo} method.
The result obtained according to the methodology required by the NEMA-NU2-2018 norm~\cite{NEMA:2018} is shown in the right panel of Fig.~\ref{fig::necr}.
The maximum value of NECR for a standard length of the phantom is equal to 630 kcps (550 kcps) and is achieved at about 30 kBq/cc (25 kBq/cc) for AFOV of 200~cm and 140~cm respectively.
The longer the phantom, the lower value of activity concentration for which the NECR peak is obtained.
%However, narrow cut may be used only to the phantom recommended in the NEMA norm, while wider cuts proposed by us may be used also for the patient (or phantom with diameter of about 50 cm).

\FloatBarrier

%==========================================================

\subsection{Sensitivity}

\begin{figure}[h]
\centering
\includegraphics[width=0.49\textwidth]{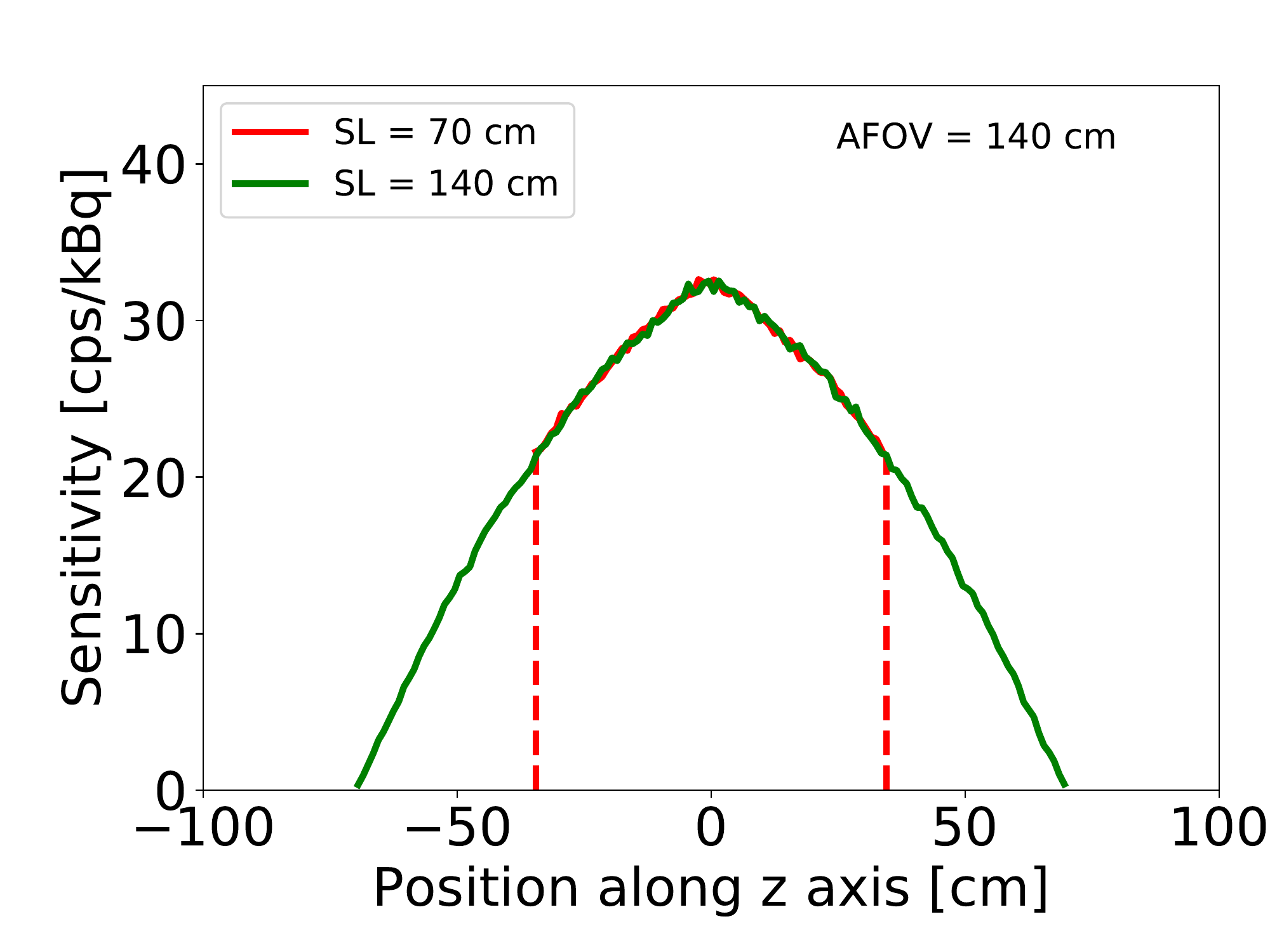}
\includegraphics[width=0.49\textwidth]{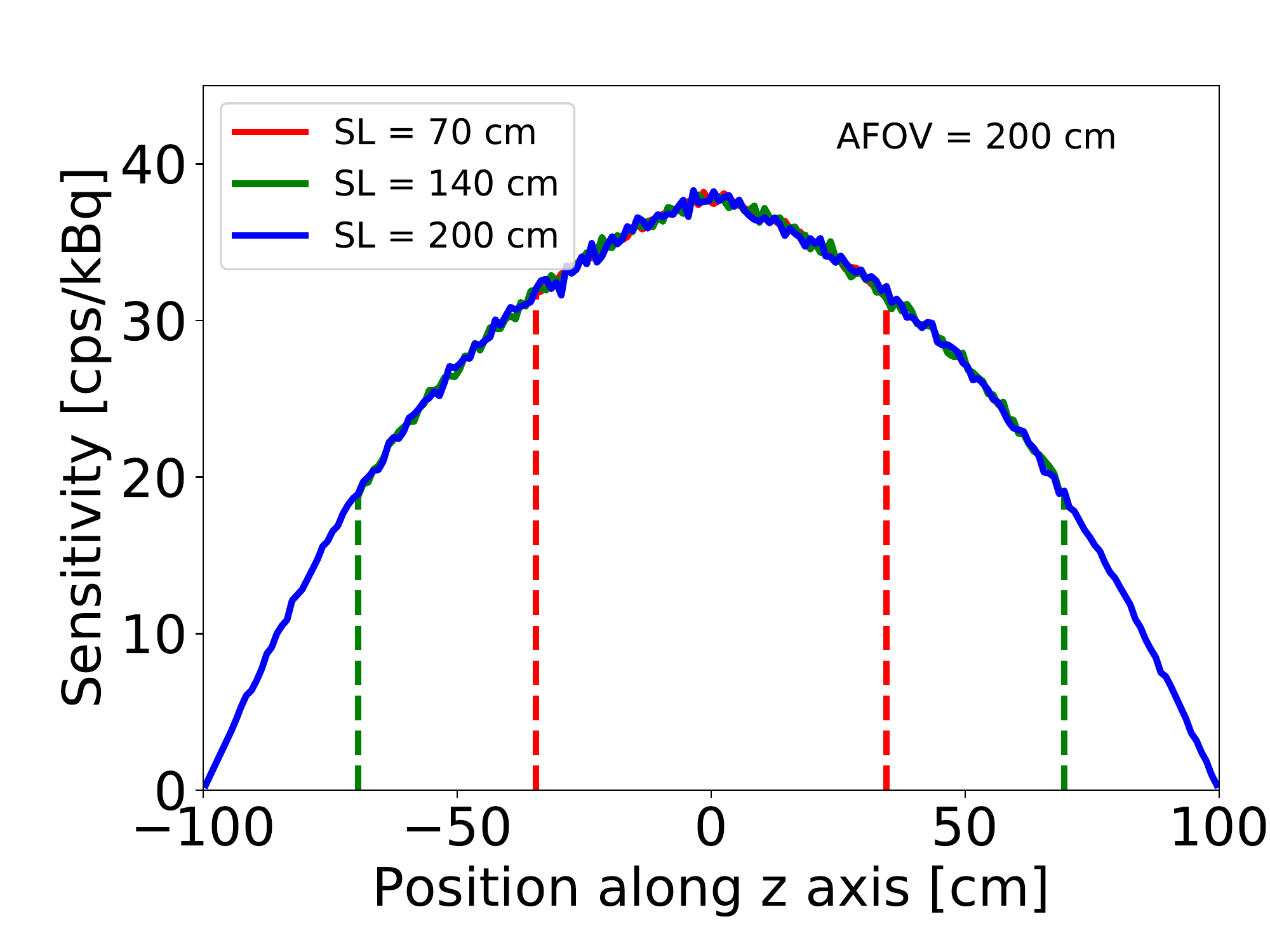}
\caption{
Sensitivity profiles for 2 values of AFOV and 3 values of source lengths (SL) as it is indicated in the legends.
}
\label{fig::sensitivity_profiles_lengths}
\end{figure}

Determined values of sensitivities are summarized in %Tab.~\ref{tab::sensitivities}
Fig.~\ref{fig::sensitivity_profiles_lengths} showing the sensitivity profiles  grouped according to the length of the detector. 
The value of sensitivity determined at the centre of the tomograph amounts to 38~cps/kBq and 32~cps/kBq for AFOV of 200~cm and 140~cm, respectively.
Results for AFOV of 140 cm and SL of 200 cm were omitted.

\FloatBarrier

%==========================================================

\subsection{Image Quality}

\begin{figure}[h]
\centering
\includegraphics[width=\textwidth]{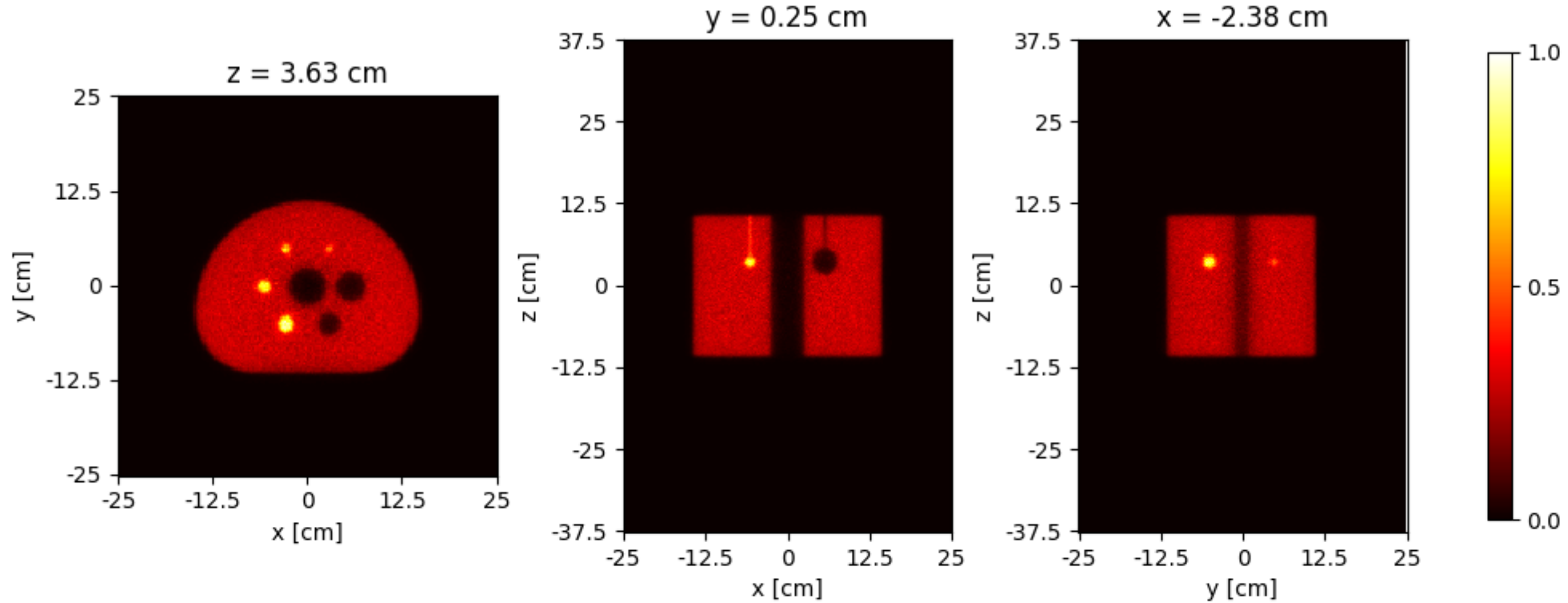}
\caption{
NEMA IEC phantom simulation for 140-cm-long J-PET scanner reconstructed with the CASTOR software. Images were obtained after 5 iteration. 
}
\label{fig::iq_reconstructed_phantom}
\end{figure}

In Fig.~\ref{fig::iq_reconstructed_phantom} exemplary reconstructed NEMA IEC images are presented for 5th iteration of TOF-MLEM in axial (left), coronal (middle) and sagittal (right) view. All the spheres could be easily distinguished from the background.

\begin{figure}[h]
\centering
\includegraphics[width=0.6\textwidth]{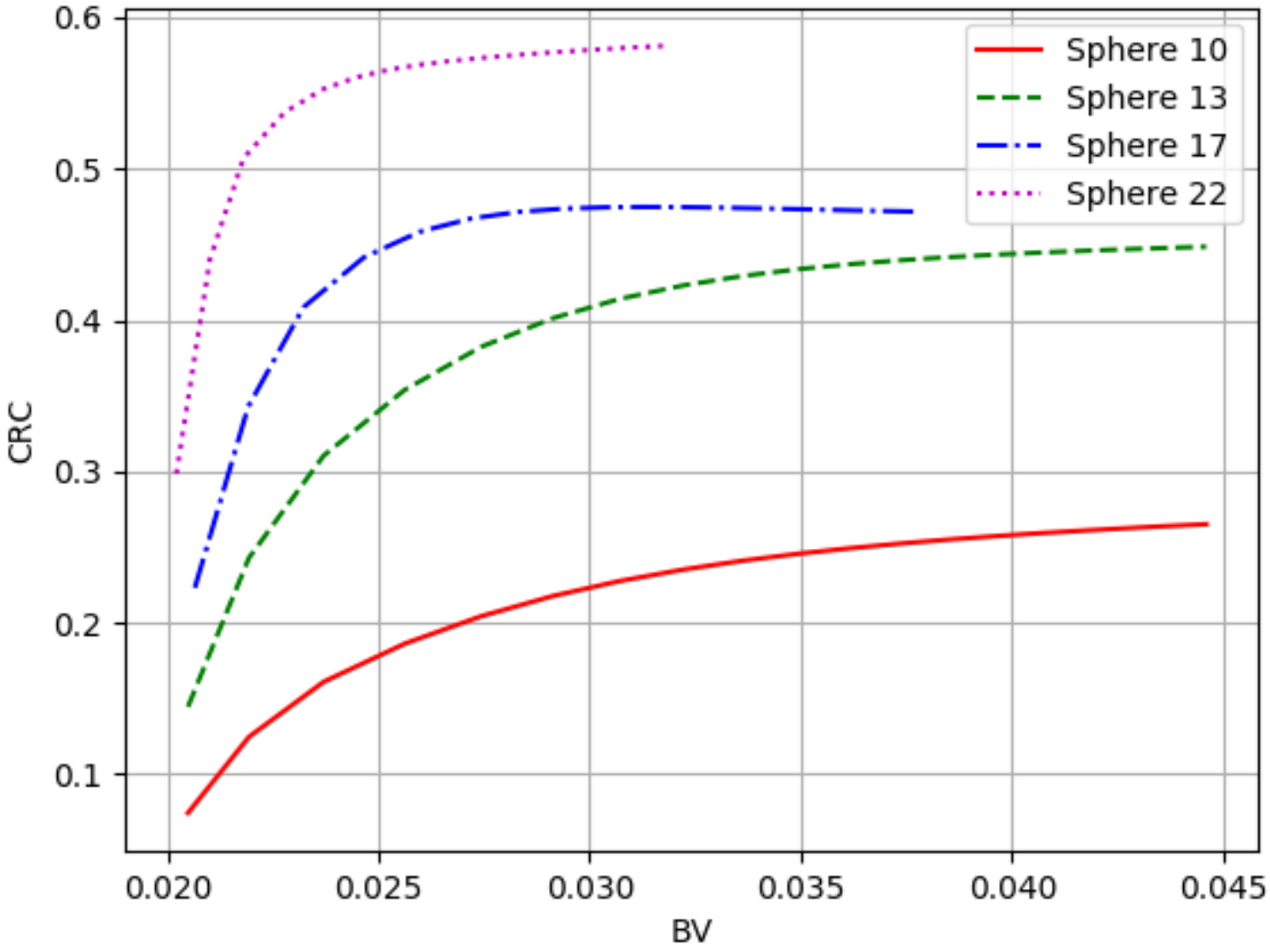}
\caption{
CRC versus BV in the reconstructed images of four hot spheres calculated for TOF-MLEM reconstruction algorithm in axial (left), coronal (middle) and sagittal (right) view.
}
\label{fig::crc_vs_bv}
\end{figure}

CRC vs BV plots for hot spheres are presented in Fig.~\ref{fig::crc_vs_bv}. All cases shown, typical trade-off between the CRC and BV parameters. The greatest value of CRC is observed for the biggest sphere (Sphere 22) and the lowest for Sphere 10. Values of the BV started from 0.02 and increased with the number of iterations.

\FloatBarrier

%==========================================================
\section{Discussion}
\label{sec::discussion}
%==========================================================

The performance characteristics presented in the above section were determined applying the prescription given in the NEMA-NU2-2018 norm~\cite{NEMA:2018}.
This norm requires estimations of SF and NECR using 70~cm long linear source and the scatter phantom of the same length.
Such length was adjusted formerly to the short AFOV PET scanners being much shorter than 70~cm.
In this article we considered total-body PET scanners with the AFOV of 140~cm and 200~cm and therefore, in addition to NEMA standards we performed estimations of the performance characteristics also for the source and phantom length of 140~cm and 200~cm.
Tab.~\ref{tab::sf_results} indicates that while increasing the length of the phantom from 70~cm to 140~cm and 200~cm, the scatter fraction increases only slightly from  35.6\% (36.2\%) to 37.4\% (37.6\%) and 38.2\% (38\%), respectively for AFOV of 140~cm (200~cm).
In turn, the maximum of NECR value (Fig.~\ref{fig::necr}) is decreasing from 550~kcps (630~kcps) to 420~kcps (540~kcps) and 320~kcps (450~kcps) when increasing the phantom and source lengths from 70~cm to 140~cm and 200~cm, respectively for AFOV of 140~cm (200~cm).

The spatial resolution was determined for the ideal case when the DOI is known, for the cases when it is reconstructed with the precision of FWHM(DOI)~=~10~mm and when it is unknown.
In general we observed that in the center of the considered TB-J-PET scanner the radial and tangential resolution is better than 4.9~mm independent of the scenario and that the axial resolution varies between PSF~=~4.8~mm (for the ideal case) and PSF~=~7.8~mm (for the unknown DOI).
In the case when DOI is determined with the uncertainty of 10~mm the radial spatial resolution varies between 4.3~mm (at the center) to 6.3~mm at the most distant NEMA defined point.
The tangential resolution is equal to about 3.9~mm for all positions and axial PSF varies between 6.4~mm and 7.6~mm.

In Tab.~\ref{tab::pet_systems} the characteristics simulated for the double layer TB-J-PET are compared with the state-of-the-art PET scanners and the first TB-PET systems.

The spatial resolution estimated for the TB-J-PET (built from strips with 6~mm times 30~mm cross section) is equal to 4.8 mm transversal and 7.6~mm axial.
Obtained results show that the spatial resolution depends significantly on the DOI resolution and can be improved down to PSF~=~3.7~mm (transversal) and 4.8~mm (axial) when improving the DOI reconstruction precision.
Thus the values of the spatial resolution expected for the TB-J-PET are comparable with the resolution of current clinical PET scanners.

The value of SF of 36.2\% obtained for the TB-J-PET is in the range of typical scatter fractions for other systems.
The NECR is by 3.5 to 5 times larger with respect to standard PET and 2.4 times lower than of uExplorer.

The sensitivity in the center of the scanner is higher by 1.7 to 3.9 times with respect to current PET systems and lower by a~factor of 4.6 and 1.4 with respect to uExplorer and PennPET Explorer, respectively.
Yet, the sensitivity of TB-J-PET may be further increased by adding a~third detection layer.

Reconstructed images of the NEMA IEC phantom and the CRC versus BV curves shows that good quality images could be obtained with the total body J-PET setup. However, the impact of other effects (point spread modelling, post-reconstruction filtering, different reconstruction algorithms, other corrections etc.) have to be further carefully investigated.

In case of the total-body scan, the figure of merit (TB-FOM), a measure of the whole-body sensitivity, can be expressed as the rate of events registered and selected for the image reconstruction per rate of photons emitted from the whole-body.
Taking into account an approximately triangular shape of the sensitivity profile, in the case of the 200~cm long source TB-FOM can be approximated as 
TB-FOM = 0.5 times S(@ 0cm) times AFOV / 200 cm.
Comparing values of TB-FOM shown in Tab.~\ref{tab::pet_systems} one can infer that the sensitivity for the total-body scan (TB-FOM) of the J-PET is by factor of 12.6 to 38 larger with respect to TB-FOM of the current clinical PET scanners (in agreement with estimations presented in articles~\cite{VandenbergheMoskal2020,Moskal2020clinical}), and it is by factor of 4.4 less than TB-FOM of uExplorer.

\begin{table}[!htb]
 \setlength\tabcolsep{3pt} % default value: 6pt
\footnotesize
\centering
\begin{tabular}{|c|c|c|c|c|c|c|c|}
\hline
  \multirow{2}{*}{Scanner} &
  AFOV &
  SF &
  NECR Peak &
  S @ 0 cm &
  TB-FOM &
  \multicolumn{2}{c|}{FWHM @ 1 cm [mm]} \\ 
\cline{7-8}
   &
  [cm] &
  [\%] &
  [kcps @ kBq/cc] &
  [cps/kBq] &
  [cps/kBq] &
  transversal &
  axial \\
\hline
  TB-J-PET (DOI = 10 mm) &
  \multirow{3}{*}{200} & % AFOV
  \multirow{3}{*}{36.2} & % SF
  \multirow{3}{*}{630 @ 30} & % NECR Peak
  \multirow{3}{*}{38} & % S @ 0 cm
  \multirow{3}{*}{19} & % TB-FOM
  4.3 & % transversal
  7.6 \\ % axial
\cline{1-1} \cline{7-8}
  TB-J-PET (DOI not known) &
   & % AFOV
   & % SF
   & % NECR Peak
   & % S @ 0 cm
   & % TB-FOM
  4.8 & % transversal
  7.8 \\ % axial
  \cline{1-1} \cline{7-8}
  TB-J-PET (DOI known) &
   & % AFOV
   & % SF
   & % NECR Peak
   & % S @ 0 cm
   & % TB-FOM
  3.7 & % transversal
  4.9 \\ % axial
  \hline
  J-PET~\cite{Kowalski2018}  &
  100 & % AFOV
  34.7 & % SF
  300 @ 40 & % NECR Peak
  14.9 & % S @ 0 cm
  3.7 & % TB-FOM
  3 & % transversal
  6 \\ % axial
\hline
  GE: %PET/CT: 
  Discovery IQ (5 rings)~\cite{GEDiscoveryIQ, Llompart2017} &
  26 & % AFOV
  36.2 & % SF
  124 @ 9.1 & % NECR Peak
  22.8 & % S @ 0 cm
  1.5 & % TB-FOM
  4.2 & % transversal
  4.2 \\ % axial
\hline
  Siemens: %PET/CT: 
  Biograph mCT~\cite{SiemensBiograph, Karlberg2016, Ghabrial2018} &
  21.8 & % AFOV
  33.2 & % SF
  180.3 @ 29 & % NECR Peak
  9.7 & % S @ 0 cm
  0.5 & % TB-FOM
  4.4 & % transversal
  4.4 \\ % axial
\hline
  Philips: %PET/CT
  Vereos~\cite{PhilipsVereos, Miller2016} &
  16.4 & % AFOV
  31.6 & % SF
  157.6 @ 52.8 & % NECR Peak
  22.1 & % S @ 0 cm
  0.9 & % TB-FOM
  3.99 & % transversal
  3.99 \\ % axial
\hline
  uExplorer~\cite{spencer2020performance} & %WAS: VandenbergheMoskal2020 
  194.0 & % AFOV
  36.3 & % SF
  1524 @ 17.3 & % NECR Peak
  174.0 & % S @ 0 cm
  84.4 & % TB-FOM
  3.0 & % transversal
  2.8 \\ % axial
\hline
  PennPET Explorer~\cite{VandenbergheMoskal2020} &
  70 & % AFOV
  32 & % SF
  $>$ 1200 & % NECR Peak
  55 & % S @ 0 cm
  9.6 & % TB-FOM
  4 & % transversal
  4 \\ % axial
\hline
\end{tabular}
\caption{
Table summarizing properties of different PET systems.
Sources and phantom used for estimating SF, NECR and sensitivity have standard length of 70 cm.
TB-J-PET is the geometry investigated in this article (double-layer with AFOV = 200 cm and the plastics strips cross-section of 6 mm times 30 mm), while the J-PET is the geometry investigated in previous studies (AFOV = 100 cm, D = 75 cm, double-layer with plastic strips with cross sections of 4 mm times 20 mm) described in detail in Ref.~\cite{Kowalski2018}.
\label{tab::pet_systems}
}
\end{table}

\section{Conclusions}
\label{sec::Conclusions}
%==========================================================

Performance characteristics of the total-body J-PET scanner (TB-J-PET) built from plastic scintillators were estimated according to the NEMA NU 2-2018 standards~\cite{NEMA:2018}. The calculations were performed by means of the GATE simulation package~\cite{Jan2004,Jan2011,sarrut2014review,sarrut2021review} and the J-PET software analysis tools~\cite{Krzemien2020,Kowalski2018}. For the TB-J-PET a double layer geometry with plastic strips with dimensions of 6~mm times 30~mm was assumed. The strips are arranged in modules forming 24-sided polygon with the inner diameter of~78.6~cm.

The performed simulations indicated that for the TB-J-PET with AFOV~=~200~cm a spatial resolution of PSF~=~3.7~mm (transversal) and PSF~=~4.9~mm (axial) is achievable. The NECR peak of 630 kcps is expected at 30 kBq/cc activity concentration, and the sensitivity at the center amounts to 38 cps/kBq. The scatter fraction is estimated to 36.2\%.  

The values of scatter fraction and spatial resolution are comparable to those obtained for the clinical PET scanners as well as to the first total-body uExplorer PET.
However, the NECR  is by factor of about 4 larger with respect to standard PET systems. Moreover, the TB-J-PET sensitivity for the whole-body scan (TB-FOM) is by a factor of 12.6 to 38 larger with respect to current clinical PET systems with AFOV in the range from 16~cm to 26~cm, and by a factor of about 4.9 less than of uExplorer PET.  

Though the whole-body sensitivity increase of the TB-J-PET is less than the one of uExplorer it is still a~significant improvement with respect to the current 16-26 cm long PET systems (and can still be increased by adding a~third detection layer).
This makes J-PET an economic and alternative technology for the construction of the total-body PET systems.
The cost of the TB-J-PET is expected to be a factor of 5 lower with respect to the LYSO based TB-PET systems~\cite{Moskal2020clinical}.
This is mainly due to the  application of the axially arranged long strips of plastic scintillators with readout at the edges, instead of the detectors built from radially arranged blocks of heavy scintillator crystals, thus reducing significantly not only the cost of scintillators but also the number of SiPMs and electronics channels.
In addition to the discussed advantages, the mechanical robustness of plastics with respect to crystals allows for making the plastic total-body scanner lightweight, modular and portable, thus making TB-J-PET a~promising cost-effective solution for the broad clinical applications of total-body PET scanners.

Sensitivity, scatter fraction and NECR were obtained not only for standard NEMA sources and phantoms with 70 cm length but also for longer ones with lengths of 140 cm and 200 cm equal to the length of scanners investigated in this article.
We propose that the NECR characteristic should be redefined and the characteristic should be presented not for activity concentration but activity concentration normalized by the length of the phantom or just for the total activity used.

%==========================================================

\section*{Acknowledgments}

The authors acknowledge technical and administrative support of A.~Heczko, M.~Kajetanowicz and W.~Migda\l{}.
This work was supported by the National Centre for Research and Development through grant INNOTECH-K1/IN1/64/159174/NCBR/12, the Foundation for Polish Science through the TEAM POIR.04.04.00-00-4204/17 programes, the National Science Centre of Poland through grants no.\ 2017/25/N/NZ1/00861, 2019/35/B/ST2/03562, the Ministry of Science and Higher Education through grants no. 6673/IA/SP/2016, 7150/E-338/SPUB/2017/1, the Austrian Science Fund through grant no.\ FWF-P26783 and the Jagiellonian University via project CRP/0641.221.2020.
B acknowledges the support of InterDokMed project no. POWR.03.02.00-00-I013/16.

\section*{Disclosure of Conflicts of Interest}
The authors have no relevant conflicts of interest to disclose.

\newcommand{\newblock}{}
\bibliographystyle{unsrtnat}
\bibliography{total_body_modular_pet}

\begin{thebibliography}{100}
\providecommand{\natexlab}[1]{#1}
\providecommand{\url}[1]{\texttt{#1}}
\expandafter\ifx\csname urlstyle\endcsname\relax
  \providecommand{\doi}[1]{doi: #1}\else
  \providecommand{\doi}{doi: \begingroup \urlstyle{rm}\Url}\fi

\bibitem[{Schmall~J.~P., Karp~J.~S. and Alavi~A.}(2019)]{schmall2019}
{Schmall~J.~P., Karp~J.~S. and Alavi~A.}
\newblock The potential role of total body pet imaging in assessment of
  atherosclerosis.
\newblock \emph{PET clinics}, 14\penalty0 (2):\penalty0 245--250, 2019.

\bibitem[{McKenney-Drake~M.~L., Moghbel~M.C., Paydary~K. et
  al.}(2018)]{mckenney2018}
{McKenney-Drake~M.~L., Moghbel~M.C., Paydary~K. et al.}
\newblock 18 f-naf and 18 f-fdg as molecular probes in the evaluation of
  atherosclerosis.
\newblock \emph{European journal of nuclear medicine and molecular imaging},
  45\penalty0 (12):\penalty0 2190--2200, 2018.

\bibitem[{Grant~A.~M., Deller~T.~W., Khalighi~M.~M. et al.}(2016)]{grant2016}
{Grant~A.~M., Deller~T.~W., Khalighi~M.~M. et al.}
\newblock Nema nu 2-2012 performance studies for the sipm-based tof-pet
  component of the ge signa pet/mr system.
\newblock \emph{Medical Physics}, 43\penalty0 (5):\penalty0 2334--2343, 2016.

\bibitem[{Van Sluis~J., De Jong~J., Schaar~J. et al.}(2019)]{vansluis2019}
{Van Sluis~J., De Jong~J., Schaar~J. et al.}
\newblock Performance characteristics of the digital biograph vision pet/ct
  system.
\newblock \emph{Journal of Nuclear Medicine}, 60\penalty0 (7):\penalty0
  1031--1036, 2019.

\bibitem[{Houshmand~S., Salavati~A., Hess~S., et al.}(2015)]{houshmand2015}
{Houshmand~S., Salavati~A., Hess~S., et al.}
\newblock An update on novel quantitative techniques in the context of evolving
  whole-body pet imaging.
\newblock \emph{PET clinics}, 10\penalty0 (1):\penalty0 45--58, 2015.

\bibitem[{Vandenberghe~S., Moskal~P. and
  Karp~J.~S.}(2020)]{VandenbergheMoskal2020}
{Vandenberghe~S., Moskal~P. and Karp~J.~S.}
\newblock State of the art in total body pet.
\newblock \emph{{EJNMMI Phys.}}, 7:\penalty0 35, 2020.

\bibitem[{Moskal~P. and
  St{\k{e}}pie{\'n}~E.~{\L{}}.}(2020)]{Moskal2020clinical}
{Moskal~P. and St{\k{e}}pie{\'n}~E.~{\L{}}.}
\newblock Prospects and clinical perspectives of total-body {PET} imaging using
  plastic scintillator.
\newblock \emph{{PET Clinics}}, 15:\penalty0 439, 2020.

\bibitem[{Badawi~R.~D., Shi~H., Hu~P. et al.}(2019)]{Badawi2019}
{Badawi~R.~D., Shi~H., Hu~P. et al.}
\newblock First human imaging studies with the explorer total-body pet.
\newblock \emph{The Journal of Nuclear Medicine}, 60\penalty0 (3):\penalty0
  299--303, 2019.

\bibitem[{Karp~J.~S., Viswanath~V., Geagan~M.~J. et al.}(2020)]{karp2020}
{Karp~J.~S., Viswanath~V., Geagan~M.~J. et al.}
\newblock Pennpet explorer: design and preliminary performance of a whole-body
  imager.
\newblock \emph{Journal of Nuclear Medicine}, 61\penalty0 (1):\penalty0
  136--143, 2020.

\bibitem[{Cherry~S.~R., Badawi~R.~D., Karp~J.~S. et al.}(2017)]{Cherry2017}
{Cherry~S.~R., Badawi~R.~D., Karp~J.~S. et al.}
\newblock Total-body imaging: Transforming the role of positron emission
  tomography.
\newblock \emph{Science Translational Medicine}, 9\penalty0 (381), 2017.
\newblock ISSN 1946-6234.
\newblock \doi{10.1126/scitranslmed.aaf6169}.

\bibitem[{Cherry~S.~R., Jones~T., Karp~J.~S. et al.}(2018)]{Cherry2018}
{Cherry~S.~R., Jones~T., Karp~J.~S. et al.}
\newblock {Total-Body PET: Maximizing Sensitivity to Create New Opportunities
  for Clinical Research and Patient Care}.
\newblock \emph{The Journal of Nuclear Medicine}, 59\penalty0 (1):\penalty0
  3--12, 2018.

\bibitem[{Zhang~X., Xie~Z., Berg~E. et al.}(2020)]{zhang2020}
{Zhang~X., Xie~Z., Berg~E. et al.}
\newblock Total-body dynamic reconstruction and parametric imaging on the
  uexplorer.
\newblock \emph{Journal of Nuclear Medicine}, 61\penalty0 (2):\penalty0
  285--291, 2020.

\bibitem[{Surti~S., Viswanath~V., Daube-Witherspoom~M.~E. et
  al.}(2020)]{surti2020}
{Surti~S., Viswanath~V., Daube-Witherspoom~M.~E. et al.}
\newblock Benefit of improved performance with state-of-the art digital pet/ct
  for lesion detection in oncology.
\newblock \emph{Journal of Nuclear Medicine}, pages jnumed--120, 2020.

\bibitem[{Surti~S., Pantel~A.~R. and Karp~J.~S.}(2020)]{surti2020total}
{Surti~S., Pantel~A.~R. and Karp~J.~S.}
\newblock Total body pet: Why, how, what for?
\newblock \emph{IEEE Transactions on Radiation and Plasma Medical Sciences},
  4\penalty0 (3):\penalty0 283--292, 2020.

\bibitem[{Jones~T.}(2020)]{jones2020total}
{Jones~T.}
\newblock Total body pet imaging from mice to humans.
\newblock \emph{Frontiers in Physics}, 8:\penalty0 77, 2020.

\bibitem[{Efthimiou~N.}(2020)]{efthimiou2020new}
{Efthimiou~N.}
\newblock New challenges for pet image reconstruction for total-body imaging.
\newblock \emph{{PET Clinics}}, 15:\penalty0 453, 2020.

\bibitem[{Majewski~S.}(2020)]{Majewski2020}
{Majewski~S.}
\newblock {Imaging is believing: The future of human total-body molecular
  imaging starts now}.
\newblock \emph{IL NUOVO CIMENTO}, 100\penalty0 (8):\penalty0 43, 2020.

\bibitem[{Nardo~L., Schmall~J.~P., Werner~T.~J. et al.}(2020)]{nardo2020}
{Nardo~L., Schmall~J.~P., Werner~T.~J. et al.}
\newblock Potential roles of total-body pet/computed tomography in pediatric
  imaging.
\newblock \emph{PET Clinics}, 15\penalty0 (3):\penalty0 271--279, 2020.

\bibitem[{Nakajima~R., Abe~K. and Sakai~S.}(2017)]{nakajima2017}
{Nakajima~R., Abe~K. and Sakai~S.}
\newblock Igg4-related diseases; whole-body fdg-pet/ct may be easier to
  evaluate rare lesions.
\newblock \emph{Journal of Nuclear Medicine}, 58\penalty0 (supplement
  1):\penalty0 943--943, 2017.

\bibitem[{Yamashita~H., Kubota~K. and Mimori~A.}(2014)]{yamashita2014}
{Yamashita~H., Kubota~K. and Mimori~A.}
\newblock Clinical value of whole-body pet/ct in patients with active rheumatic
  diseases.
\newblock \emph{Arthritis research \& therapy}, 16\penalty0 (4):\penalty0 423,
  2014.

\bibitem[{Borja~A.~J., Rojulpote~C., Hancin~E.~C. et
  al.}(2020)]{borja2020update}
{Borja~A.~J., Rojulpote~C., Hancin~E.~C. et al.}
\newblock An update on the role of total-body pet imaging in the evaluation of
  atherosclerosis.
\newblock \emph{PET Clinics}, 2020.

\bibitem[{Zhang~X., Xie~Z., Berg~E. et al.}(2019)]{zhang2019}
{Zhang~X., Xie~Z., Berg~E. et al.}
\newblock Total-body parametric imaging using kernel and direct reconstruction
  on the uexplorer.
\newblock \emph{Journal of Nuclear Medicine}, 60\penalty0 (supplement
  1):\penalty0 456--456, 2019.

\bibitem[{Surti~S., Werner~M.~E. and Karp~J.~S.}(2013)]{surti2013}
{Surti~S., Werner~M.~E. and Karp~J.~S.}
\newblock Study of pet scanner designs using clinical metrics to optimize the
  scanner axial fov and crystal thickness.
\newblock \emph{Physics in Medicine and Biology}, 58\penalty0 (12):\penalty0
  3995, 2013.

\bibitem[{Zhang~J., Knopp~M.~I. and Knopp~M.~V.}(2019)]{zhang2019sparse}
{Zhang~J., Knopp~M.~I. and Knopp~M.~V.}
\newblock Sparse detector configuration in sipm digital photon counting pet: a
  feasibility study.
\newblock \emph{Molecular Imaging and Biology}, 21\penalty0 (3):\penalty0
  447--453, 2019.

\bibitem[{Zein~S.~A., Karakatsanis~N.~A., Issa~M. et
  al.}(2020)]{zein2020physical}
{Zein~S.~A., Karakatsanis~N.~A., Issa~M. et al.}
\newblock Physical performance of a long axial field-of-view pet scanner
  prototype with sparse rings configuration: A monte carlo simulation study.
\newblock \emph{Medical Physics}, 47\penalty0 (4):\penalty0 1949--1957, 2020.

\bibitem[{Zhang~Y. and Wong~W.-H.}(2017)]{zhang2017system}
{Zhang~Y. and Wong~W.-H.}
\newblock System design studies for a low-cost high-resolution bgo pet with
  1-meter axial field of view.
\newblock \emph{Journal of Nuclear Medicine}, 58\penalty0 (supplement
  1):\penalty0 221--221, 2017.

\bibitem[{Gonzalez-Montoro~A., Sanchez~F., Majewski~S. et
  al.}(2017)]{gonzalez2017highly}
{Gonzalez-Montoro~A., Sanchez~F., Majewski~S. et al.}
\newblock Highly improved operation of monolithic bgo-pet blocks.
\newblock \emph{Journal of Instrumentation}, 12\penalty0 (11):\penalty0 C11027,
  2017.

\bibitem[{Brunner~S.~E. and Schaart~D.~R.}(2017)]{brunner2017bgo}
{Brunner~S.~E. and Schaart~D.~R.}
\newblock Bgo as a hybrid scintillator/cherenkov radiator for cost-effective
  time-of-flight pet.
\newblock \emph{Physics in Medicine and Biology}, 62\penalty0 (11):\penalty0
  4421, 2017.

\bibitem[{Kwon~S.~I., Roncali~E., Gola~A. et al.}(2019)]{kwon2019dual}
{Kwon~S.~I., Roncali~E., Gola~A. et al.}
\newblock Dual-ended readout of bismuth germanate to improve timing resolution
  in time-of-flight pet.
\newblock \emph{Physics in Medicine and Biology}, 64\penalty0 (10):\penalty0
  105007, 2019.

\bibitem[{Cates~J.~W. and Levin~C.~S.}(2019)]{cates2019}
{Cates~J.~W. and Levin~C.~S.}
\newblock Electronics method to advance the coincidence time resolution with
  bismuth germanate.
\newblock \emph{Physics in Medicine and Biology}, 64\penalty0 (17):\penalty0
  175016, 2019.

\bibitem[{Gundacker~S., Turtos~R.~M., Kratochwil~N., et
  al.}(2020)]{gundacker2020experimental}
{Gundacker~S., Turtos~R.~M., Kratochwil~N., et al.}
\newblock Experimental time resolution limits of modern sipms and tof-pet
  detectors exploring different scintillators and cherenkov emission.
\newblock \emph{Physics in Medicine and Biology}, 65\penalty0 (2):\penalty0
  025001, 2020.

\bibitem[{Grignon~C., Barbet~J., Bardi{\`e}s~M. et al.}(2007)]{Grignon2007}
{Grignon~C., Barbet~J., Bardi{\`e}s~M. et al.}
\newblock Nuclear medical imaging using $\beta$+ $\gamma$ coincidences from
  {44Sc} radio-nuclide with liquid xenon as detection medium.
\newblock \emph{Nuclear Instruments and Methods in Physics Research Section A:
  Accelerators, Spectrometers, Detectors and Associated Equipment},
  571\penalty0 (1-2):\penalty0 142--145, 2007.

\bibitem[{Donnard~J., Chen~W.-T., Cussonneau J.-P. et al.}(2012)]{Donnard2012}
{Donnard~J., Chen~W.-T., Cussonneau J.-P. et al.}
\newblock Compton imaging with liquid xenon and {44Sc}: recent progress toward
  3 gamma imaging.
\newblock \emph{Nuclear Medicine Review}, 15\penalty0 (C):\penalty0 64--67,
  2012.

\bibitem[{Lang~C., Habs~D., Thirolf~P.~G. and Zoglauer~A.}(2012)]{Lang2012}
{Lang~C., Habs~D., Thirolf~P.~G. and Zoglauer~A.}
\newblock Submillimeter nuclear medical imaging with a {Compton Camera} using
  triple coincidences of collinear$\backslash$beta+ annihilation photons
  and$\backslash$gamma-rays.
\newblock \emph{arXiv preprint arXiv:1202.0397}, 2012.

\bibitem[{Lang~C., Habs~D., Parodi~K. and Thirolf~P.~G.}(2014)]{Lang2014}
{Lang~C., Habs~D., Parodi~K. and Thirolf~P.~G.}
\newblock Sub-millimeter nuclear medical imaging with high sensitivity in
  positron emission tomography using $\beta$+ $\gamma$ coincidences.
\newblock \emph{Journal of Instrumentation}, 9\penalty0 (01):\penalty0 P01008,
  2014.

\bibitem[{Oger~T., Chen~W.-T., Cussonneau~J.-P. et al.}(2012)]{Oger2012}
{Oger~T., Chen~W.-T., Cussonneau~J.-P. et al.}
\newblock A liquid xenon {TPC} for a medical imaging {Compton} telescope.
\newblock \emph{Nuclear Instruments and Methods in Physics Research Section A:
  Accelerators, Spectrometers, Detectors and Associated Equipment},
  695:\penalty0 125--128, 2012.

\bibitem[{Thirolf~P.~G., Lang~C. and Parodi~K.}(2015)]{Thirolf2015}
{Thirolf~P.~G., Lang~C. and Parodi~K.}
\newblock Perspectives for highly-sensitive pet-based medical imaging using
  $\beta$+ $\gamma$ coincidences.
\newblock \emph{Acta Physica Polonica, A.}, 127\penalty0 (5), 2015.

\bibitem[{Yoshida~E., Tashima~H., Nagatsu~K. et al.}(2020)]{Yoshida2020}
{Yoshida~E., Tashima~H., Nagatsu~K. et al.}
\newblock {Whole gamma imaging: a new concept of PET combined with Compton
  imaging}.
\newblock \emph{Physics in Medicine and Biology}, 65\penalty0 (12):\penalty0
  125013, 2020.

\bibitem[Kuramoto et~al.(2017)Kuramoto, Nakamori, Kimura, Gunji, Takakura, and
  Kataoka]{kuramoto2017development}
M~Kuramoto, T~Nakamori, S~Kimura, S~Gunji, M~Takakura, and J~Kataoka.
\newblock Development of tof-pet using compton scattering by plastic
  scintillators.
\newblock \emph{Nuclear Instruments and Methods in Physics Research Section A:
  Accelerators, Spectrometers, Detectors and Associated Equipment},
  845:\penalty0 668--672, 2017.

\bibitem[Shimazoe et~al.(2020)Shimazoe, Yoshino, Ohshima, Uenomachi, Oogane,
  Orita, Takahashi, Kamada, Yoshikawa, and Takahashi]{shimazoe2020development}
Kenji Shimazoe, Masao Yoshino, Yusuke Ohshima, Mizuki Uenomachi, Kenichiro
  Oogane, Tadashi Orita, Hiroyuki Takahashi, Kei Kamada, Akira Yoshikawa, and
  Miwako Takahashi.
\newblock Development of simultaneous pet and compton imaging using gagg-sipm
  based pixel detectors.
\newblock \emph{Nuclear Instruments and Methods in Physics Research Section A:
  Accelerators, Spectrometers, Detectors and Associated Equipment},
  954:\penalty0 161499, 2020.

\bibitem[Hemmati et~al.(2017)Hemmati, Kamali-Asl, Ay, and
  Ghafarian]{hemmati2017compton}
Hamidreza Hemmati, Alireza Kamali-Asl, Mohammadreza Ay, and Pardis Ghafarian.
\newblock Compton scatter tomography in tof-pet.
\newblock \emph{Physics in Medicine \& Biology}, 62\penalty0 (19):\penalty0
  7641, 2017.

\bibitem[Kishimoto et~al.(2017)Kishimoto, Kataoka, Taya, Tagawa, Mochizuki,
  Ohsuka, Nagao, Kurita, Yamaguchi, Kawachi, et~al.]{kishimoto2017first}
Aya Kishimoto, Jun Kataoka, Takanori Taya, Leo Tagawa, Saku Mochizuki, Shinji
  Ohsuka, Yuto Nagao, Keisuke Kurita, Mitsutaka Yamaguchi, Naoki Kawachi,
  et~al.
\newblock First demonstration of multi-color 3-d in vivo imaging using
  ultra-compact compton camera.
\newblock \emph{Scientific reports}, 7\penalty0 (1):\penalty0 1--7, 2017.

\bibitem[Uenomachi et~al.(2020)Uenomachi, Mizumachi, Yoshihara, Takahashi,
  Shimazoe, Yabu, Yoneda, Watanabe, Takeda, Orita, et~al.]{uenomachi2020double}
Mizuki Uenomachi, Yuki Mizumachi, Yuri Yoshihara, Hiroyuki Takahashi, Kenji
  Shimazoe, Goro Yabu, Hiroki Yoneda, Shin Watanabe, Shin’ichiro Takeda,
  Tadashi Orita, et~al.
\newblock Double photon emission coincidence imaging with gagg-sipm compton
  camera.
\newblock \emph{Nuclear Instruments and Methods in Physics Research Section A:
  Accelerators, Spectrometers, Detectors and Associated Equipment},
  954:\penalty0 161682, 2020.

\bibitem[{Moskal~P., Salabura~P., Silarski~M. et al.}(2011)]{Moskal2011}
{Moskal~P., Salabura~P., Silarski~M. et al.}
\newblock Novel detector systems for the positron emission tomography.
\newblock \emph{Bio-Algorithms and Med-Systems}, 7:\penalty0 73--78, 2011.

\bibitem[{Moskal~P., Nied{\'z}wiecki~S., Bednarski~T. et
  al.}(2014)]{Moskal:2014sra}
{Moskal~P., Nied{\'z}wiecki~S., Bednarski~T. et al.}
\newblock {Test of a single module of the J-PET scanner based on plastic
  scintillators}.
\newblock \emph{Nucl. Instrum. Meth.}, A764:\penalty0 317--321, 2014.
\newblock \doi{10.1016/j.nima.2014.07.052}.

\bibitem[{Moskal~P., Zo{\'n}~N., Bednarski~T. et al.}(2015)]{Moskal:2014rja}
{Moskal~P., Zo{\'n}~N., Bednarski~T. et al.}
\newblock {A novel method for the line-of-response and time-of-flight
  reconstruction in TOF-PET detectors based on a library of synchronized model
  signals}.
\newblock \emph{Nucl. Instrum. Meth.}, A775:\penalty0 54--62, 2015.
\newblock \doi{10.1016/j.nima.2014.12.005}.

\bibitem[{Moskal~P., Rundel~O., Alfs~D. et al.}(2016)]{Moskal:2016ztv}
{Moskal~P., Rundel~O., Alfs~D. et al.}
\newblock {Time resolution of the plastic scintillator strips with matrix
  photomultiplier readout for J-PET tomograph}.
\newblock \emph{Physics in Medicine and Biology}, 61:\penalty0 2025, 2016.
\newblock \doi{10.1088/0031-9155/61/5/2025}.

\bibitem[{Nied{\'z}wiecki~S., Bia{\l}as~P., Curceanu~C. et
  al.}(2017)]{Szymon-Acta}
{Nied{\'z}wiecki~S., Bia{\l}as~P., Curceanu~C. et al.}
\newblock {J-PET: a new technology for the whole-body PET imaging}.
\newblock \emph{Acta Phys. Polon.}, B48:\penalty0 1567, 2017.
\newblock \doi{10.5506/APhysPolB.48.1567}.

\bibitem[{Sharma~N.~G., Silarski~M., Chhokar~J et al.}(2020)]{sharma2020hit}
{Sharma~N.~G., Silarski~M., Chhokar~J et al.}
\newblock Hit-time and hit-position reconstruction in strips of plastic
  scintillators using multi-threshold readouts.
\newblock \emph{IEEE Transactions on Radiation and Plasma Medical Sciences},
  4:\penalty0 528, 2020.

\bibitem[{Sharma~S., Chhokar~J., Curceanu~C. et
  al.}(2020)]{sharma2020estimating}
{Sharma~S., Chhokar~J., Curceanu~C. et al.}
\newblock {Estimating relationship between the Time Over Threshold and energy
  loss by photons in plastic scintillators used in the {J-PET} scanner}.
\newblock \emph{EJNMMI physics}, 7\penalty0 (1):\penalty0 1--15, 2020.

\bibitem[{Mao~R., Wu~C., Lu~S. et al.}(2013)]{mao2013crystal}
{Mao~R., Wu~C., Lu~S. et al.}
\newblock Crystal growth and scintillation properties of lso and lyso crystals.
\newblock \emph{Journal of crystal growth}, 368:\penalty0 97--100, 2013.

\bibitem[{Saint-Gobain Crystals}(2018{\natexlab{a}})]{saintgobain_lyso}
{Saint-Gobain Crystals}.
\newblock {LYSO Scintillation Material Data Sheet}, 2018{\natexlab{a}}.

\bibitem[{National Institute of Standards and Technology}()]{nist}
{National Institute of Standards and Technology}.
\newblock URL \url{https://www.nist.gov/pml}.
\newblock Accessed 15 Jul 2020.

\bibitem[{Saint-Gobain Crystals}()]{saintgobain}
{Saint-Gobain Crystals}.
\newblock URL \url{https://www.crystals.saint-gobain.com}.

\bibitem[{Vilardi~I., Braem~A., Chesi~E. et
  al.}(2006)]{vilardi2006optimization}
{Vilardi~I., Braem~A., Chesi~E. et al.}
\newblock Optimization of the effective light attenuation length of yap: Ce and
  lyso: Ce crystals for a novel geometrical pet concept.
\newblock \emph{Nuclear Instruments and Methods in Physics Research Section A:
  Accelerators, Spectrometers, Detectors and Associated Equipment},
  564\penalty0 (1):\penalty0 506--514, 2006.

\bibitem[{Mao~R., Zhang~L. and Zhu~R.-Y.}(2008)]{mao2008optical}
{Mao~R., Zhang~L. and Zhu~R.-Y.}
\newblock Optical and scintillation properties of inorganic scintillators in
  high energy physics.
\newblock \emph{IEEE Transactions on Nuclear Science}, 55\penalty0
  (4):\penalty0 2425--2431, 2008.

\bibitem[{Saint-Gobain Crystals}(2016)]{saintgobain_bgo}
{Saint-Gobain Crystals}.
\newblock {BGO Scintillation Material Data Sheet}, 2016.

\bibitem[{Chen~J., Zhang~L. and Zhu~R.-Y.}(2004)]{chen2004large}
{Chen~J., Zhang~L. and Zhu~R.-Y.}
\newblock Large size {LYSO} crystals for future high energy physics
  experiments.
\newblock In \emph{IEEE Symposium Conference Record Nuclear Science 2004.},
  volume~1, pages 117--125. IEEE, 2004.

\bibitem[{Saint-Gobain Crystals}(2018{\natexlab{b}})]{saintgobain_bc}
{Saint-Gobain Crystals}.
\newblock {BC-400, BC-404, BC-408, BC-412, BC-416 Plastic Scintillators Data
  Sheet}, 2018{\natexlab{b}}.

\bibitem[{Moskal~P., Jasi{\'n}ska~B., St{\k{e}}pie{\'n}~E. and
  Bass~S.~D.}(2019)]{moskal2019positronium}
{Moskal~P., Jasi{\'n}ska~B., St{\k{e}}pie{\'n}~E. and Bass~S.~D.}
\newblock Positronium in medicine and biology.
\newblock \emph{Nature Reviews Physics}, 1\penalty0 (9):\penalty0 527--529,
  2019.

\bibitem[{Moskal~P., Kisielewska~D., Curceanu~C. et
  al.}(2019)]{moskal2019feasibility}
{Moskal~P., Kisielewska~D., Curceanu~C. et al.}
\newblock {Feasibility study of the positronium imaging with the J-PET
  tomograph}.
\newblock \emph{{Physics in Medicine and Biology}}, 64\penalty0 (5):\penalty0
  055017, 2019.

\bibitem[{Moskal~P., Alfs~D., Bednarski~T. et al.}(2016)]{moskal2016potential}
{Moskal~P., Alfs~D., Bednarski~T. et al.}
\newblock Potential of the {J-PET} detector for studies of discrete symmetries
  in decays of positronium atom-a purely leptonic system.
\newblock \emph{Acta Phys. Polon.}, B47:\penalty0 509, 2016.

\bibitem[{Gajos~A.}(2020)]{gajos2020sensitivity}
{Gajos~A.}
\newblock Sensitivity of discrete symmetry tests in the positronium system with
  the {J-PET} detector.
\newblock \emph{Symmetry}, 12\penalty0 (8):\penalty0 1268, 2020.

\bibitem[{Hiesmayr~B.~C. and Moskal~P.}(2019)]{hiesmayr2019witnessing}
{Hiesmayr~B.~C. and Moskal~P.}
\newblock {Witnessing entanglement in Compton scattering processes via mutually
  unbiased bases}.
\newblock \emph{Scientific reports}, 9\penalty0 (1):\penalty0 1--14, 2019.

\bibitem[{Moskal~P., Krawczyk~N., Hiesmayr~B.~C. et
  al.}(2018)]{moskal2018feasibility}
{Moskal~P., Krawczyk~N., Hiesmayr~B.~C. et al.}
\newblock Feasibility studies of the polarization of photons beyond the optical
  wavelength regime with the {J-PET} detector.
\newblock \emph{The European Physical Journal C}, 78\penalty0 (11):\penalty0
  970, 2018.

\bibitem[{Kami{\'n}ska~D., Gajos~A., Czerwi{\'n}ski~E. et
  al.}(2016)]{kaminska2016feasibility}
{Kami{\'n}ska~D., Gajos~A., Czerwi{\'n}ski~E. et al.}
\newblock A feasibility study of ortho-positronium decays measurement with the
  {J-PET} scanner based on plastic scintillators.
\newblock \emph{The European Physical Journal C}, 76\penalty0 (8):\penalty0
  1--14, 2016.

\bibitem[NEMA(2018)]{NEMA:2018}
NEMA.
\newblock Nema standards publication nu 2-2018: Performance measurements of
  positron emission tomographs (pet), 2018.
\newblock National Electrical Manufacturers Association (NEMA NU 2-2018).

\bibitem[{Jan~S., Santin~G., Strul~D. et al.}(2004)]{Jan2004}
{Jan~S., Santin~G., Strul~D. et al.}
\newblock {GATE}: a simulation toolkit for {PET} and {SPECT}.
\newblock \emph{Physics in Medicie and Biology}, 49:\penalty0 4543--4561, 2004.

\bibitem[{Jan~S., Benoit~D., Becheva~E. et al.}(2011)]{Jan2011}
{Jan~S., Benoit~D., Becheva~E. et al.}
\newblock {GATE} v6: a major enhancement of the {GATE} simulation platform
  enabling modelling of {CT} and radiotherapy.
\newblock \emph{Physics in Medicie and Biology}, 56:\penalty0 881--901, 2011.

\bibitem[{Sarrut~D., Bardi{\`e}s~M., Boussion~N. et
  al.}(2014)]{sarrut2014review}
{Sarrut~D., Bardi{\`e}s~M., Boussion~N. et al.}
\newblock {A review of the use and potential of the GATE Monte Carlo simulation
  code for radiation therapy and dosimetry applications}.
\newblock \emph{Medical physics}, 41\penalty0 (6Part1):\penalty0 064301, 2014.

\bibitem[{Sarrut~D., Bala~M., Bardies~M. et al.}(2021)]{sarrut2021review}
{Sarrut~D., Bala~M., Bardies~M. et al.}
\newblock {Advanced Monte Carlo simulations of emission tomography imaging
  systems with GATE}.
\newblock \emph{Physics in Medicine and Biology}, 66\penalty0 (10TR03), 2021.
\newblock \doi{10.1088/1361-6560/abf276}.

\bibitem[{Kowalski~P., Wi{\'s}licki~W., Shopa~R.~Y. et
  al.}(2018)]{Kowalski2018}
{Kowalski~P., Wi{\'s}licki~W., Shopa~R.~Y. et al.}
\newblock {Estimating the NEMA characteristics of the J-PET tomograph using the
  GATE package}.
\newblock \emph{Physics in Medicine and Biology}, 63:\penalty0 165008, 2018.

\bibitem[{Krzemie{\'n}~W., Gajos~A., Kacprzak~K. et al.}(2020)]{Krzemien2020}
{Krzemie{\'n}~W., Gajos~A., Kacprzak~K. et al.}
\newblock {J-PET Framework: Software platform for PET tomography data
  reconstruction and analysis}.
\newblock \emph{SoftwareX}, 11:\penalty0 100487, 2020.

\bibitem[{Zhang~X., Zhou~J., Cherry~S.~R. et al.}(2017)]{Zhang2017}
{Zhang~X., Zhou~J., Cherry~S.~R. et al.}
\newblock {Quantitative image reconstruction for total-body PET imaging using
  the 2-meter long EXPLORER scanner}.
\newblock \emph{Physics in Medicine and Biology}, 62\penalty0 (6):\penalty0
  2465, 2017.
\newblock URL \url{http://stacks.iop.org/0031-9155/62/i=6/a=2465}.

\bibitem[{Eljen Technology}()]{eljentechnology}
{Eljen Technology}.
\newblock Physical constants of plastic scintillators.
\newblock URL
  \url{https://eljentechnology.com/images/technical_library/Physical_Constants_Plastic.pdf}.

\bibitem[{Conti~M.}(2009)]{Conti2009}
{Conti~M.}
\newblock State of the art and challenges of time-of-flight pet.
\newblock \emph{Physica Medica}, 25:\penalty0 1--11, 2009.

\bibitem[{Smyrski~J., Moskal~P., Bednarski~T. et
  al.}(2014)]{smyrski2014application}
{Smyrski~J., Moskal~P., Bednarski~T. et al.}
\newblock Application of wls strips for position determination in strip pet
  tomograph based on plastic scintillators.
\newblock \emph{Bio-Algorithms and Med-Systems}, 10\penalty0 (2):\penalty0
  59--63, 2014.

\bibitem[{Smyrski~J., Alfs~D., Bednarski~T. et al.}(2017)]{Smyrski2017}
{Smyrski~J., Alfs~D., Bednarski~T. et al.}
\newblock {Measurement of gamma quantum interaction point in plastic
  scintillator with WLS strips}.
\newblock \emph{Nucl. Instrum. Meth.}, A851:\penalty0 39--42, 2017.
\newblock \doi{10.1016/j.nima.2017.01.045}.

\bibitem[{Shivani~S., {\L}uczy{\'n}ska~E., Heinze~S. and
  Moskal~P.}(2020)]{shivani2020development}
{Shivani~S., {\L}uczy{\'n}ska~E., Heinze~S. and Moskal~P.}
\newblock Development of j-pem for breast cancer detection.
\newblock \emph{Acta Physica Polonica. A}, 137\penalty0 (2), 2020.

\bibitem[{Spencer~B.~A., Berg~E., Schmall~J. et
  al.}(2020)]{spencer2020performance}
{Spencer~B.~A., Berg~E., Schmall~J. et al.}
\newblock {Performance evaluation of the uEXPLORER Total-body PET/CT scanner
  based on NEMA NU 2-2018 with additional tests to characterize long axial
  field-of-view PET scanners}.
\newblock \emph{Journal of Nuclear Medicine}, pages jnumed--120, 2020.

\bibitem[{Kowalski~P., Wi{\'s}licki~W., Raczy{\'n}ski~L. et
  al.}(2016)]{Kowalski2016}
{Kowalski~P., Wi{\'s}licki~W., Raczy{\'n}ski~L. et al.}
\newblock Scatter fraction of the j-pet tomography scanner.
\newblock \emph{Acta Phys. Polon}, B47:\penalty0 549, 2016.

\bibitem[{Kowalski~P., Moskal~P., Wi{\'s}licki~W. et al.}(2015)]{Kowalski2015}
{Kowalski~P., Moskal~P., Wi{\'s}licki~W. et al.}
\newblock Multiple scattering and accidental coincidences in the j-pet detector
  simulated using gate package.
\newblock \emph{Acta Phys. Polon}, A127:\penalty0 1505--1512, 2015.

\bibitem[{Thielemans~K., Tsoumpas~C., Mustafovic~S. et al.}(2012)]{stir}
{Thielemans~K., Tsoumpas~C., Mustafovic~S. et al.}
\newblock {STIR: software for tomographic image reconstruction release 2}.
\newblock \emph{Physics in Medicine and Biology}, 57\penalty0 (4):\penalty0
  867, 2012.

\bibitem[{Khateri~P., Fischer~J., Lustermann~W. et
  al.}(2019)]{khateri2019implementation}
{Khateri~P., Fischer~J., Lustermann~W. et al.}
\newblock Implementation of cylindrical pet scanners with block detector
  geometry in stir.
\newblock \emph{EJNMMI physics}, 6\penalty0 (1):\penalty0 15, 2019.

\bibitem[{Shopa~R.~Y., Klimaszewski~K., Kowalski~P. et al.}(2017)]{Shopa2017}
{Shopa~R.~Y., Klimaszewski~K., Kowalski~P. et al.}
\newblock Three-dimensional image reconstruction in j-pet using filtered
  back-projection method.
\newblock \emph{Acta Phys. Polon. B}, 48:\penalty0 1757, 2017.

\bibitem[{Moskal~P., Kisielewska~D., Shopa~R.~Y. et
  al.}(2020)]{moskal2020performance}
{Moskal~P., Kisielewska~D., Shopa~R.~Y. et al.}
\newblock Performance assessment of the 2 $\gamma$ positronium imaging with the
  total-body pet scanners.
\newblock \emph{{EJNMMI Phys.}}, 7:\penalty0 44, 2020.

\bibitem[{Conti~M., Bendriem~B., Casey~M. et al.}(2005)]{Conti2005}
{Conti~M., Bendriem~B., Casey~M. et al.}
\newblock First experimental results of time-of-flight reconstruction on an lso
  pet scanner.
\newblock \emph{Physics in Medicine and Biology}, 50\penalty0 (19):\penalty0
  4507, 2005.

\bibitem[{Bailey~D.~L., Townsend~D.~W., Valk~P.~E. and
  Maisey~M.~N.}(2005)]{PETBasicScience}
{Bailey~D.~L., Townsend~D.~W., Valk~P.~E. and Maisey~M.~N.}
\newblock \emph{Positron Emission Tomography, Basic Sciences}.
\newblock Springer, 2005.

\bibitem[{Shopa~R.~Y.}(2020)]{shopa2020estimation}
{Shopa~R.~Y.}
\newblock Estimation of spatial resolution for 3-layer j-pet scanner using tof
  fbp based on event-by-event approach.
\newblock \emph{Acta Physica Polonica B}, 51\penalty0 (1), 2020.

\bibitem[{Becker~R., Buck~A., Casella~C. et al.}(2017)]{BECKER2017}
{Becker~R., Buck~A., Casella~C. et al.}
\newblock {The SAFIR experiment: Concept, status and perspectives}.
\newblock \emph{{Nuclear Instruments and Methods in Physics Research Section A:
  Accelerators, Spectrometers, Detectors and Associated Equipment}},
  845:\penalty0 648--651, 2017.

\bibitem[Merlin et~al.(2018)Merlin, Stute, Benoit, Bert, Carlier, Comtat,
  Filipovic, Lamare, and Visvikis]{merlin2018castor}
Thibaut Merlin, Simon Stute, Didier Benoit, Julien Bert, Thomas Carlier, Claude
  Comtat, Marina Filipovic, Fr{\'e}d{\'e}ric Lamare, and Dimitris Visvikis.
\newblock Castor: a generic data organization and processing code framework for
  multi-modal and multi-dimensional tomographic reconstruction.
\newblock \emph{Physics in Medicine and Biology}, 63\penalty0 (18):\penalty0
  185005, 2018.

\bibitem[{Moskal~P. and Smyrski~J.}(2018)]{patentWLS}
{Moskal~P. and Smyrski~J.}
\newblock Detecting device for determining a position of reaction of gamma
  quanta and a method for determining a position of reaction of a gamma quanta
  in positron emission tomography, August~7 2018.
\newblock US Patent 10,042,058.

\bibitem[{Yang~X., Peng H.}(2015)]{Yang2015}
{Yang~X., Peng H.}
\newblock The use of noise equivalent count rate and the {NEMA} phantom for
  {PET} image quality evaluation.
\newblock \emph{Physica Medica}, 31:\penalty0 179--184, 2015.

\bibitem[{GE}()]{GEDiscoveryIQ}
{GE}.
\newblock Producer's website with technical details.
\newblock URL
  \url{http://www3.gehealthcare.com/en/products/categories/molecular_imaging/pet-ct/pet-ct_scanners/discovery_iq}.
\newblock 2018-03-29.

\bibitem[{Reyn\'es-Llompart G., G{\'a}mez-Cenzano~C., Romero-Zayas~I. et
  al.}(2017)]{Llompart2017}
{Reyn\'es-Llompart G., G{\'a}mez-Cenzano~C., Romero-Zayas~I. et al.}
\newblock Performance characteristics of the whole-body {D}iscovery {IQ}
  {PET/CT} system.
\newblock \emph{The Journal of Nuclear Medicine}, 58\penalty0 (7):\penalty0
  1155--1161, 2017.

\bibitem[{Siemens}()]{SiemensBiograph}
{Siemens}.
\newblock Online catalogue with technical details.
\newblock URL \url{http://www.activexray.com/pdf/Siemens\_Biograph.pdf}.
\newblock 2018-03-29.

\bibitem[{Karlberg~A.~M., S{\ae}ther~O., Eikenes~L. et
  al.}(2016)]{Karlberg2016}
{Karlberg~A.~M., S{\ae}ther~O., Eikenes~L. et al.}
\newblock Quantitative comparison of {PET} performance - {S}iemens {B}iograph
  m{CT} and m{MR}.
\newblock \emph{European Journal of Nuclear Medicine and Molecular Imaging
  Physics}, 3:\penalty0 5, 2016.

\bibitem[{Ghabrial~A., Franklin~D. and Zaidi~H.}(2018)]{Ghabrial2018}
{Ghabrial~A., Franklin~D. and Zaidi~H.}
\newblock A {M}onte {C}arlo simulation study of the impact of novel
  scintillation crystals on performance characteristics of {PET} scanners.
\newblock \emph{Physica Medica}, 50:\penalty0 37--45, 2018.

\bibitem[{Philips}()]{PhilipsVereos}
{Philips}.
\newblock Philips {V}ereos {PET/CT} brochure.
\newblock URL
  \url{https://philipsproductcontent.blob.core.windows.net/assets/20170523/360753349c5d4a6aa46ba77c015e75b4.pdf}.
\newblock 2020-06-29.

\bibitem[{Miller~M.~A.}(2016)]{Miller2016}
{Miller~M.~A.}
\newblock Focusing on high performance.
\newblock \emph{Philips brochure}, 2016.
\newblock URL \url{www.philips.com/VereosPETCT}.

\end{thebibliography}

\end{document}